\documentclass[a4paper,fleqn]{cas-sc}

\usepackage{amssymb,amsmath}
\usepackage{relsize}                % mathlarger
\usepackage{color}                  % textcolor
\usepackage{gensymb}                % degree
\usepackage{float}                  % H in figure position
\usepackage{subcaption}             % subtable
\usepackage{changepage}             % adjustwidth
\usepackage{multirow}               % multirow
\usepackage{enumitem}               % enumerate
\usepackage{longtable}              % LTleft, LTright
\usepackage{adjustbox}              % adjustbox
\usepackage[utf8]{inputenc}
\usepackage{collcell}
\usepackage{pgf}
\def\colorModel{rgb} %You can use rgb or hsb
\def\cca#1{
    \pgfmathsetmacro\celVal{#1}
    \pgfmathsetmacro\compA{\celVal>=0?0.9:0.6}      %Component R or H
    \pgfmathsetmacro\compB{\celVal>=0?0.6:0.9}       %Component G or S
    \pgfmathsetmacro\compC{0.6}      %Component B or B
    \edef\clrmacro{\noexpand\cellcolor[\colorModel]{\compA,\compB,\compC}}\clrmacro
    \ifdim \compA pt =1pt\color{white}\fi{#1}
}

\usepackage{geometry}           % To reduce the margins

\usepackage{tikz}
\usetikzlibrary{shapes,positioning}

\usepackage{algorithm}
\usepackage{algpseudocode}

\usepackage{lineno, hyperref}
\modulolinenumbers[5]

\usepackage[authoryear]{natbib}
\def\tsc#1{\csdef{#1}{\textsc{\lowercase{#1}}\xspace}}
\tsc{WGM}
\tsc{QE}
\begin{document}
\let\WriteBookmarks\relax
\def\floatpagepagefraction{1}
\def\textpagefraction{.001}

\shorttitle{Data Processing Framework for Ship Performance Analysis}    

\shortauthors{P. Gupta, Y.-R. Kim, S. Steen, A. Rasheed}  

\title[mode = title]{Data Processing Framework for Ship Performance Analysis}  

% \tnotemark[<tnote number>] 



\author[1]{Prateek Gupta}[orcid=0000-0001-7147-0868]

\cormark[1]


\ead{prateek.gupta@ntnu.no}

% \ead[url]{<URL>}

\credit{Conceptualization, Methodology, Investigation, Software, Data curation, Visualization, Writing}

\affiliation[1]{organization={Norwegian University of Science and Technology (NTNU)},
            addressline={Department of Marine Technology}, 
            city={Trondheim},
%          citysep={}, % Uncomment if no comma needed between city and postcode
            postcode={7052}, 
            state={Sør-trondelag},
            country={Norway}}

\author[1]{Young-Rong Kim}

% Footnote of the second author

% Email id of the second author
\ead{youngrong.kim@ntnu.no}

% \ead[url]{<URL>}

\credit{Conceptualization, Methodology, Investigation, Software, Data curation, Visualization, Writing}

\author[1]{Sverre Steen}%[<options>]

% Footnote of the second author
% \fnmark[3]

% Email id of the second author
\ead{sverre.steen@ntnu.no}


\credit{Conceptualization, Supervision, Resources}

\author[2]{Adil Rasheed}%[<options>]

% Footnote of the second author

% Email id of the second author
\ead{adil.rasheed@ntnu.no}


\credit{Conceptualization, Supervision}

\affiliation[2]{organization={Norwegian University of Science and Technology (NTNU)},
            addressline={Department of Engineering Cybernetics}, 
            city={Trondheim},
%          citysep={}, % Uncomment if no comma needed between city and postcode
            postcode={7034}, 
            state={Sør-trondelag},
            country={Norway}}

\cortext[1]{Corresponding author}



\begin{abstract}
The hydrodynamic performance of a sea-going ship can be analysed using the data obtained from the ship. Such data can be gathered from different sources, like onboard recorded in-service data, AIS data, and noon reports. Each of these sources is known to have their inherent problems. The current work gives a brief introduction to these data sources as well as the common problems associated with them, along with some examples. In order to resolve most of these problems, a streamlined semi-automatic data processing framework for fast data processing is developed and presented here. The data processing framework can be used to process the data obtained from any of the above three mentioned sources. The framework incorporates processing steps like interpolating weather hindcast (metocean) data to ship's location in time, deriving additional features, validating data, estimating resistance components, data cleaning, and outlier detection. A brief description of each of the processing steps is provided with examples from existing datasets. The processed data can be further used to analyse the hydrodynamic performance of a ship.
\end{abstract}


\begin{highlights}
\item Ship's hydrodynamic performance can be assessed using the data from ship-in-service 
\item Data can be recorded onboard the ship, obtained via AIS or as noon reports 
\item Gathered data needs to be cleaned and processed for performance analysis
\item A standardized data processing framework for preparing the data is developed
\item Data processing framework can be easily casted for processing any similar datasets
\end{highlights}

% Keywords
% Each keyword is separated by \sep

\maketitle 

% \linenumbers

\section{Introduction}

The performance of a sea-going ship is important not only to keep the fuel and operational cost in-check but also to reduce global emissions from the shipping industry. Analyzing the performance of a ship is also of great interest for charter parties to estimate the potential of a ship and the profit that can be made out of it. Therefore, driven by both the economic and social incentives, the trade of ship performance analysis and monitoring has been booming substantially in recent times. The importance of in-service data in this context is very well understood by most of the stake holders, clearly reflected by the amount of investment made by them on onboard sensors, data acquisition systems, and onshore operational performance monitoring and control centers.

The traditional way to evaluate the performance of a ship is using the noon report data provided by the ship's crew. A more exact approach, but not very feasible for commercial vessels, was suggested by \citet{Walker2007}, conducting in-service sea trials in calm-water conditions on a regular basis. With the advent of sensor-based continuous monitoring systems, the current trend is to directly or indirectly observe the evolution of the calm-water speed-power curve over time. ISO 19030 \cite{ISO19030} along with several researchers (\citet{Koboevic2019}; \citet{Coraddu2019DigTwin}) recommends observing the horizontal shift (along the speed axis) of the calm-water speed-power curve, termed as the speed-loss, over time to monitor the performance of a sea-going ship using the in-service data. Alternatively, it is suggested to observe the vertical shift of the calm-water speed-power curve, often termed as the change in power demand (adopted by \citet{Gupta2021PrefMon} and \citet{CARCHEN2020}). Some researchers also formulated and used some indirect performance indicators like fuel consumption (\citet{Koboevic2019}), resistance (or fouling) coefficient (\citet{Munk2006}; \citet{Foteinos2017}; \citet{CARCHEN2020}), (generalized) admiralty coefficient (\citet{Ejdfors2019}; \citet{Gupta2021}), wake fraction (\citet{CARCHEN2020}), fuel efficiency (\citet{Kim2021}), etc. In each of these cases, it is clearly seen (and most of the time acknowledged) that the results are quite sensitive to the quality of the data used to estimate the ship's performance.
% customary measure -> power demand or fuel consumption at a fixed speed, near-calm water using noon ?

% How the data quality affects the ship performance analysis (brief explanation, citation)
% (From ABS) Marine and offshore environments and operations have unique variables such as noise, dust, temperature, humidity, electronic and magnetic interference, and location since they are typically far from land-based infrastructures. This has a significant impact on the performance, reliability, and longevity of the sensors, cables, and data communication and storage devices. Data collected, transmitted, and stored in these severe environments and operations may potentially have significant quality issues, such as data loss, invalid values, transmission delay, incorrect timestamp order, among others.
% -> Data quality for IoT data is highly dependent on the data processing quality

% Some prior studies related to ship data processing (ex, Oyvind...)
% -> Most of them deal with specific data and specific problem cases.
% ->  (Motivation of this paper) 
% Above contents should be connected with the previous and the next paragraphs.

The ship's performance-related data obtained from various sources usually inherits some irregularities due to several factors like sensor inaccuracies, vibration of the sensor mountings, electrical noise, variation of environment, etc., as pointed out in the Guide for Smart Functions for Marine Vessels and Offshore Units (Smart Guide) published recently by \citet{ABS2020guide}. The quality of data used to carry-out ship performance analysis and the results obtained further can be significantly improved by adopting some rational data processing techniques, as shown by \citet{Liu2020} and \citet{Kim2020}. Another important factor is the source of data as it may also be possible to obtain such datasets using the publicly available AIS data (\citet{You2017}). \citet{Dalheim2020DataPrep} presented a data preparation toolkit based on the in-service data recorded onboard 2 ships. The presented toolkit was developed for a specific type of dataset, where the variables were recorded asynchronously and had to be synchronized before carrying-out ship performance analysis. The current work would rather focus on challenges faced while processing an already synchronized dataset.

The current paper presents a review of different data sources used for ship performance analysis and monitoring, namely, onboard recorded in-service data, AIS data, and noon reports, along with the characteristics for each of these data sources. Finally, a data processing framework is outlined which can be used to prepare these datasets for ship performance analysis and monitoring. Although the data processing framework is developed for the performance monitoring of ships, it may easily be casted for several other purposes. With the easy availability of data from ships, the concept of creating digital twins for sea-going ships is becoming quite popular. \citet{Major2021} presented the concept of digital twin for a ship and the cranes onboard it. The digital twin established by \citet{Major2021} can be used to perform three main offshore operations, including remote monitoring of the ship, maneuvering in harsh weather and crane operations, from an onshore control center. Moreover, as pointed out by \citet{Major2021}, the digital twin technology can also be adopted for several other purposes, like predictive maintenance, ship autonomy, etc. Nevertheless, the data processing framework presented here can also be used to process the real-time data obtained to create digital twins for ships in an efficient manner. 

The following section discusses the art of ship performance analysis and the bare minimum characteristics of a dataset required to do such an analysis. Section \ref{sec:dataSources} presents the above mentioned sources of data used for ship performance analysis, their characteristics, and the tools required to process these datasets. Section \ref{sec:results} presents the data processing framework which can be used to process and prepare these datasets for ship performance monitoring. Finally, section \ref{sec:conclusion} finishes the paper with concluding remarks.
% The paper proposes a holistic method for processing these data sources with the main purpose of ship performance analysis. 

\section{Ship Performance Analysis}

The performance of a ship-in-service can be assessed by observing its current performance and, then, comparing it to a benchmarking standard. There are several ways to establish (or obtain) a benchmarking standard, like model test experiments, full-scale sea trials, CFD analysis, etc. It may even be possible to establish a benchmarking standard using the in-service data recorded onboard a newly built ship, as suggested by \citet{Coraddu2019DigTwin} and \citet{Gupta2021}. On the other hand, evaluating the current performance of a ship requires a good amount of data processing as the raw data collected during various voyages of a ship is susceptible to noise and errors. Moreover, the benchmarking standard is, generally, established for only a given environmental condition, most likely the calm-water condition. In order to draw a comparison between the current performance and the benchmarking standard, the current performance must be translated to the same environmental condition, therefore, increasing the complexity of the problem.

\subsection{Bare Minimum Variables}

For translating the current performance data to the benchmarking standard's environmental condition and carrying-out a reliable ship performance analysis, a list of bare minimum variables must be recorded (or observed) at a good enough sampling rate. The bare minimum list of variables must provide the following information about each sampling instant for the ship: (a) Operational control, (b) Loading condition, (c) Operational environment, and (d) Operating point. The variables containing the above information must either be directly recorded (or observed) onboard the ship, collected from regulatory data sources such as AIS, or may be derived using additional data sources, like the operational environment can be easily derived using the ship's location and timestamp with the help of an appropriate weather hindcast (or metocean) data repository.

The operational control information should contain the values of the propulsion-related control parameters set by the ship's captain on the bridge, like shaft rpm, rudder angle, propeller pitch, etc. The shaft rpm (or propeller pitch, in case of ships equipped with controllable pitch propellers running at constant rpm) is by far the most important variable here as it directly correlates with the ship's speed-through-water. It should be noted that even in the case of constant power or speed mode, the shaft rpm (or propeller pitch) continues to be the primary control parameter as the set power or speed is actually achieved by using a real-time optimizer (incorporated in the governor) which optimizes the shaft rpm (or propeller pitch) to get to the set power or speed. Nevertheless, in case the shaft rpm (or propeller pitch) is not available, it may be appropriate to use the ship's speed-through-water as an operational control parameter, as done by several researchers (\citet{FARAG2020}; \citet{Laurie2021}; \citet{Minoura2020}; \citet{Liang2019}), but in this case, it should be kept in mind that, unlike the shaft rpm (or propeller pitch), the speed-through-water is a dependant variable strongly influenced by the loading condition and the operational environment.  

The loading condition should contain the information regarding the ship's fore and aft draft, which can be easily recorded onboard the ship. Although the wetted surface area and under-water hull-form are more appropriate for a hydrodynamic analysis, these can be derived easily using the ship's hull form, if the fore and aft draft is known. The operational environment should at least contain variables indicating the intensity of wind and wave loads acting on the ship, like wind speed and direction, significant wave height, mean wave direction, mean wave period, etc. Finally, the operating point should contain the information regarding the speed-power operating point for the sampling instant. Table \ref{tab:bareMinVars} presents the list of bare minimum variables required for ship performance analysis. The list given in the table may have to be modified according to ship specifications, for example, the propeller pitch is only relevant for a ship equipped with a controllable pitch propeller.  

\begin{table}[ht]
\caption{The list of bare minimum data variables required for ship performance analysis.} \label{tab:bareMinVars}
\centering
\begin{tabular}{l|l}
\hline
\multicolumn{1}{c|}{\textbf{Category}} & \multicolumn{1}{c}{\textbf{Variables}} \\
\hline
Operational Control & Shaft rpm, Rudder angle, \& Propeller pitch \\
\hline
Loading Condition & Fore and aft draft \\
\hline
Operational Environment & \begin{tabular}[l]{@{}l@{}}Longitudinal and transverse wind speed, Significant wave height,\\ Relative mean wave direction, \& Mean wave period\end{tabular} \\
\hline
Operating Point & Shaft power \& Speed-through-water \\
\hline
\end{tabular}
\end{table}

\subsection{Best Practices} \label{sec:bestPractices}

It is well-known that the accuracy of various measurements is not the same. It also depends on the source of the measurements. The measurements recorded using onboard sensors are generally more reliable as compared to the manually recorded noon report measurements, due to the possibility of human error in the latter. Even in the case of onboard recorded sensor measurements, the accuracy varies from sensor-to-sensor and case-to-case. Some sensors can be inherently faulty, whereas others can give incorrect measurements due to unfavorable installation and operational conditions, and even the best ones are known to have some measurement noise. Thus, it is recommended to establish and follow some best practices for a reliable and robust ship performance analysis.

The onboard measurements for shaft rpm ($n$) and shaft torque ($\tau$) are generally obtained using a torsion meter installed on the propeller shaft, which is considered to be quite reliable. The shaft power ($P_s$) measurements are also derived from the same as the shaft power is related to the shaft rpm and torque through the following identity: $P_s = 2\pi n\tau$. It should be noted that no approximation is assumed in this formulation and, therefore, it should be validated with the data, if all three variables ($n, \tau, P_s$) are available. On the other hand, the measurements for speed-through-water are known to have several problems, as presented by \citet{DALHEIM2021}. Thus, it is recommended to use shaft rpm (and not speed-though-water) as the independent variable while creating data-driven regression models to predict the shaft power. Owing to the same reason, it may also be a good idea to quantify the change in ship's performance in terms of change in power demand rather than speed-loss (or speed-gain), as recommended by ISO 19030 \cite{ISO19030}.

Further, it is also quite common to use fuel oil consumption as a key performance indicator for ship performance analysis (\citet{Karagiannidis2021}). The fuel oil consumption can be easily calculated from the engine delivered torque and engine rpm, if the specific fuel consumption (SFC) curve for the engine is known. Even though the SFC curve is established and supplied by the engine manufacturer, it is only valid for a specific operating environment, and it is known to evolve over time due to engine degradation and maintenance. Thus, including the fuel oil consumption in ship performance analysis increases the complexity of the problem, which requires taking engine health into account. If the objective of ship performance analysis is also to take into account the engine performance, then it may be beneficial to divide the problem into two parts: (a) Evaluate the change in power demand (for hydrodynamic performance analysis), and (b) Evaluate the change in engine SFC (for engine performance analysis). Now, the latter can be formulated as an independent problem with a completely new set of variables-of-interest, like engine delivered torque, engine rpm, ambient air temperature, calorific value of fuel, turbocharger health, etc. This would not only improve the accuracy of ship's hydrodynamic performance analysis but would also allow the user to develop a more comprehensive and, probably, accurate analysis model. The current work is focused on the hydrodynamic performance analysis.

\subsection{Sampling Frequency}
% Thus, the final measurements obtained using these sensors are, in fact, statistical means of several measurements (known as a `sample' in statistics) obtained over a very short period of time (milliseconds).
Almost all electronics-based sensors are known to have some noise in their measurements. The simplest way adopted to subdue this noise is by taking an average over a number of measurements (known as a `sample' in statistics), recorded over a very short period of time (milliseconds). It is also known that the statistical mean of a `sample' converges to the true mean (i.e., the mean of the entire population), thereby eliminating the noise, as the number of measurements in the `sample' is increased, provided the observations follow a symmetrical distribution. Nevertheless, it is observed that the high frequency data still retains some noise, probably due to the fact that the number of measurements in each `sample' is small, i.e., the measurements are obtained by averaging a small number of samples recorded over a very short period of time. On the other hand, as seen in the case of noon reports and most of the in-service datasets, time-averaging the measurements over a longer period of time obscures the effect of moderately varying influential factors, for example, instantaneous incident wind and waves, response motions, etc. Thus, a very high sampling frequency data may retain high noise, and a very low sampling frequency data, with time-averaged values, may result in obscuring important effects from the data time-series. Furthermore, in the third scenario, it may be possible that the data acquisition (DAQ) system onboard the ship is simply using low sampling frequency, recording instantaneous values instead of time-averaged ones, saving a good amount of storage and bandwidth while transmitting it to the shore-based control centers. These low frequency instantaneous values may result in an even more degraded data quality as it would contain noise as well as obscure the moderately varying effects.

The ideal sampling frequency would also depend on the objective of the analysis and the recorded variables. For example, if the objective of the analysis is to predict the motion response of a ship or analyse its seakeeping characteristics, the data should be recorded at a high enough sampling frequency such that it is able to capture such effects. \citet{hansen2011performance} analyzed the ship's rudder movement and the resulting resistance, and demonstrated that if the sampling interval would be large, the overall dynamics of the rudder movement would not be captured, resulting in a difference in resistance. One criterion for selecting the data sampling rate is Nyquist frequency (\citet{jerri1977shannon}), which is widely used in signal processing. According to this criterion, the sampling frequency shall be more than twice the frequency of the observed phenomenon to sufficiently capture the information regarding the phenomenon. Therefore, if the aim is not to record any information regarding the above mentioned moderately varying effects (instantaneous incident wind and waves, response motions, etc.), it may be acceptable to just obtain low frequency time-averaged values so that such effects are subdued. But it may still be useful to obtain high frequency data, in this case, as it can be advantageous from data cleaning point of view. For example, the legs of time-series showing very high variance, due to the noise or moderately varying effects, can be removed from the analysis to increase the reliability of results. 

\section{Data Sources, Characteristics \& Processing Tools} \label{sec:dataSources}

\subsection{In-service Data}

The in-service data, referred to here, is recorded onboard a ship during its voyages. This is achieved by installing various sensors onboard the ship, collecting the measurements from these sensors on a regular basis (at a predefined sampling rate) using a data acquisition (DAQ) system, and fetching the collected data to onshore control centers. The two most important features of in-service data is the sampling rate (or, alternatively, sampling frequency) and the list of recorded variables. Unfortunately, there is no proper guide or standard which is followed while defining both these features for a ship. Thus, the in-service data processing has to be adapted to each case individually. 

The in-service datasets used here are recorded over a uniform (across all recorded variables) and evenly-spaced sampling interval, which makes it easier to adopt and apply data processing techniques. In an otherwise case, where the data is sampled with a non-uniform and uneven sampling interval, some more pre-processing has to be done in order to prepare it for further analysis, as demonstrated by \citet{Dalheim2020DataPrep}. \citet{Dalheim2020DataPrep} presented a detailed algorithm to deal with time vector jumps and synchronizing non-uniformly recorded data variables. The problem of synchronization can, alternatively, be looked at using the well-known dynamic time warping (DTW) technique, which is generally used for aligning the measurements taken by two sensors, measuring the same or highly correlated features. In a different approach, \citet{virtanen2020scipy} demonstrated that the collected data can be down-sampled or up-sampled (resampling) to obtain a uniform and evenly sampled dataset. 

\subsubsection{Inherently Faulty \& Incorrect Measurements} \label{sec:incorrMeasureInServData}

Some of the sensors onboard a ship can be inherently faulty and provide incorrect measurements due to unfavorable installation or operational conditions. Many of these can actually be fixed quite easily. For instance, \citet{Wahl2019} presented the case of faulty installation of the wind anemometer onboard a ship, resulting in missing measurements for head-wind condition probably due to the presence of an obstacle right in front of the sensor. Such a fault is fairly simple to deal with, say, by fixing the installation of the sensor, and it is even possible to fix the already recorded dataset using the wind measurements from one of the publicly available weather hindcast datasets. Such an instance also reflects the importance of data exploration and validation for ship performance analysis. Unlike above, the case of draft and speed-through-water measurement sensors is not as fortunate and easy to resolve.

The ship's draft is, generally, recorded using a pressure transducer installed onboard the ship. The pressure transducer measures the hydrostatic pressure acting on the bottom plate of the ship which is further converted into the corresponding water level height or the draft measurement. When the ship starts to move and the layer of water in contact with the ship develops a relative velocity with respect to the ship, the total pressure at the ship's bottom reduces due to the non-zero negative hydrodynamic pressure and, therefore, further measurements taken by the draft sensor are incorrect. This is known as the Venturi effect. It may seem like a simple case, and one may argue that the measurements can be fixed by just adding the water level height equivalent to the hydrodynamic pressure, which may be calculated using the ship's speed-through-water. Here, it should be noted that, firstly, to accurately calculate the hydrodynamic pressure, one would need the localized relative velocity of the flow (and not the ship's speed-through-water), which is impractical to measure, and secondly, the speed-though-water measurements are also known to have several sources of inaccuracy. Alternatively, it may be possible to obtain the correct draft measurements from the ship's loading computer. The loading computer can calculate the draft and trim in real-time based on the information such as the ship's lightweight, cargo weight and distribution, and ballast water loading configuration.

The state-of-the-art speed-though-water measurement device uses the Doppler acoustic speed log principle. Here, the relative speed of water around the hull (i.e., the speed-though-water) is measured by observing the shift in frequency (popularly known as the Doppler shift) of the ultrasound pulses emitted from the ship's hull, due to its motion. The ultrasonic pulses are reflected by the ocean bottom, impurities in the surrounding water, marine life, and even the liquid-liquid interface between the density difference layers in deep ocean. The speed of water surrounding the ship is influenced by the boundary layer around the hull so it is required that the ultrasonic pulses reflected only by the particles outside the boundary layer are used to estimate the speed-though-water. Therefore, a minimum pulse travelling distance has to be prescribed for the sensor. If the prescribed distance is too larger or if the ship is sailing in shallow waters, the Doppler shift is calculated using the reflection from the ocean bottom, i.e., the sensor is in ground-tracking mode, and therefore, it would clearly record the ship's speed-over-ground instead of the speed-though-water. \citet{DALHEIM2021} presented a detailed account regarding the uncertainty in the speed-though-water measurements for a ship, commenting that the speed log sensors are considered to be one of the most inaccurate ones onboard the ship. 

It may also be possible to estimate the speed-though-water of a ship using the ship's speed-over-ground and incident longitudinal water current speed. The speed-over-ground of a ship is measured using a GPS sensor, which is considered to be quite accurate, but unfortunately, the water current speed is seldom recorded onboard the ship. It is certainly possible to obtain the water current speed from a weather hindcast data source, but the hindcast measurements are not accurate enough to obtain a good estimate for speed-through-water, as indicated by \citet{Antola2017}. It should also be noted that the temporal and spatial resolution of weather hindcast data is relatively larger than the sampling interval of the data recorded onboard the ship. Moreover, the water current speed varies along the depth of the sea, therefore, the incident longitudinal water current speed must be calculated as an integral of the water current speed profile over the depth of the ship. Thus, in order to obtain accurate estimates of speed-though-water, the water current speed has to be measured or estimated upto a certain depth of the sea with good enough accuracy, which is not possible with the current state-of-the-art.

\subsubsection{Outliers} \label{sec:outliers}

Another big challenge with data processing is the problem of detecting and handling outliers. As suggested by \citet{Olofsson2020}, it may be possible to categorize outliers into the following two broad categories: (a) Contextual outliers, and (b) Correlation-defying outliers\footnote{Called collective outliers by \citet{Olofsson2020}.}. \citet{Dalheim2020DataPrep} presented methods to detect and remove contextual outliers, further categorized as (i) obvious (or invalid) outliers, (ii) repeated values, (iii) drop-outs, and (iv) spikes. Contextual outliers are easily identifiable as they either violate the known validity limits of one or more recorded variables (as seen in the case of obvious outliers and spikes) or present an easily identifiable but anomalous pattern (as seen in the case of repeated values and drop-outs). 

The case of correlation-defying outliers is much more difficult to handle, as they can easily blend into the cleaned data pool. The two most popular methods which can be used to identify correlation-defying outliers are Principal Component Analysis (PCA) and autoencoders. Both these methods try to reconstruct the data samples after learning the correlation between the variables. It is quite obvious that a correlation-defying outlier would result in an abnormally high reconstruction error and, therefore, can be detected using such techniques. In a recent attempt, \citet{Thomas2021} demonstrated an ensemble method combining PCA and autoencoders coupled with isolation forest to detect such outliers.

\subsubsection{Time-Averaging Problem} \label{sec:timeAvgProb}

As aforementioned, the onboard recorded in-service data can be supplied as time-averaged values over a short period of time (generally upto around 15 minutes). Although the time-averaging method eliminates white noise and reduces the variability in the data samples, it introduces a new problem in case of angular measurements. The angular measurements are, generally, recorded in the range of 0 to 360 degrees. When the measurement is around 0 or 360 degrees, it is obvious that the instantaneous measurements, reported by the sensor, will fluctuate in the vicinity of 0 and 360 degrees. For instance, assuming that the sensor reports a value of about 0 degree for half of the averaging time and about 360 degrees for the remaining time, the time-averaged value recorded by the data acquisition (DAQ) system will be around 180 degrees, which is significantly incorrect. Most of the angular measurements recorded onboard a ship, like relative wind direction, ship heading, etc., are known to inherit this problem, and it should be noted that, unlike the example given here, the incorrect time-averaged angle can take any value between 0 and 360 degrees, depending on the instantaneous values over which the average is calculated.  

Although it may be possible to fix these incorrect values using a carefully designed algorithm, there is no established method available at the moment. Thus, it is suggested to fix these measurements using an alternate source for the data variables. For example, the wind direction can be gathered easily from a weather hindcast data source. Thus, it can be used to correct or just replace the relative wind direction measurements, recorded onboard the ship. The ship's heading, on the other hand, can be estimated using the latitude and longitude measurements from the GPS sensor.

\subsection{AIS Data}
AIS is an automatic tracking system that uses transceivers to help ships and maritime authorities identify and monitor ship movements. It is generally used as a tool for ship transportation services to prevent collisions during navigation. Ships over 300 tons must be equipped with transponders capable of transmitting and receiving all message types of AIS under the SOLAS Convention. AIS data is divided into dynamic (position, course, speed, etc.) static (ship name, dimensions, etc.), and voyage-related data (draft, destination, ETA, etc.). Dynamic data is automatically transmitted every 2-10 seconds depending on the speed and course of the ship, and if anchored, such information is automatically transmitted every 6 minutes. On the other hand, static and voyage-related data is provided by the ship's crew, and it is transmitted every 6 minutes regardless of the ship's movement state.

Since dynamic information is automatically updated based on sensor data, it is susceptible to faults and errors, similar to those described in section \ref{sec:incorrMeasureInServData}. In addition, problems may occur even in the process of collecting and transmitting data between AIS stations, as noted by \citet{weng2020exploring}. The AIS signal can also be influenced by external factors, such as weather conditions and Earth's magnetic field, due to their interference with the very high frequency (VHF) equipment. Therefore, some of the AIS messages are lost or get mixed. Moreover, the receiving station has a short time slot during which the data must be received, and due to heavy traffic in the region, it fails to receive the data from all the ships in that time. In some cases, small ships deliver inaccurate information due to incorrectly calibrated transmitters, as shown by \citet{weng2020exploring}. In a case study, \citet{harati2007automatic} observed that 2\% of the MMSI (Maritime Mobile Service Identity) information was incorrect and 30\% of the ships were not properly marked with the correct navigation status. In the case of ship dimensions, about 18\% of the information was found to be inaccurate. Therefore, before using AIS raw data for ship performance analysis, it is necessary to check key parameters such as GPS position, speed, and course, and the data identified as incorrect must be fixed.

% \textbf{Position data with unavailable coordinates (Figures will be added)}
% \begin{itemize}
%     \item Replace with linear interpolation according to time between connected coordinates.
%     \item Positions reported over land. -> Remove data
%     \item Positions impulsive outliers -> Find outliers by using Hampel filter (peak finder), Find outliers by calculating speed between each time stamp (distance/time) \& Check with the normal speed range
% \end{itemize}

% \textbf{Incorrect draft data (Figures will be added)}
% \begin{itemize}
%     \item Check with the navigation status data (if the ship is sailing at a certain speed while the status is 0 or 5, the static data which is manually entered values have low reliability.
%     \item If the design draft of the ship is known, check the draft ratio and detect outliers through the statistical analysis. (IQR, ZSC) 
%     \item Remove or replace with rational value (ballast or laden voyage according to port information) 
% \end{itemize}

\subsubsection{Irrational Speed Data}
The GPS speed (or speed-over-ground) measurements from AIS data may contain samples that have a sudden jump compared to adjacent samples or excessively higher or lower value than the normal operating range. This type of inaccurate data can be identified through comparison with location and speed data of adjacent samples. The distance covered by the ship at the corresponding speed during the time between the two adjacent AIS messages is calculated, and the distance between the actual two coordinates is calculated using the Haversine formula (given by equation \ref{eq:havsineDistance}) to compare the two values. If the difference between the two values is negligible, the GPS speed can be said to be normal, but if not, it is recommended to be replaced with the GPS speed value of the adjacent sample. It should be noted that if the time difference between the samples is too short, the deviation of the distance calculated through this method may be large. In such a case, it is necessary to consider the average trend for several samples. If there are no valid samples nearby or the GPS coordinate data is problematic, one can refer to the normal service speed according to the ship type, as shown in table \ref{tab:vParams}, or if available, a more specific method such as normalcy box (\citet{rhodes2005maritime,tu2017exploiting}), which defines the speed range of the ships according to the geographic location, may be applied.

\begin{equation}\label{eq:havsineDistance}
{D = 2r\sin^{-1} \left(\sqrt{\sin^{2}\left(\frac{y_{i+1}-y_{i}}{2}\right)+\cos{\left(y_i\right)}\cos{\left(y_{i+1}\right)}\sin^{2}\left(\frac{x_{i+1}-x_{i}}{2}\right)}\right)}
\end{equation}

Where $D$ is the distance between two coordinates ($x_i$, $y_i$) and ($x_{i+1}$, $y_{i+1}$), $r$ is the radius of Earth, and ($x_i$, $y_i$) is the longitude and latitude at timestamp $i$.

\begin{table}[ht]
\caption{Typical service speed range of different ship types, given by \citet{solutions2018basic}.} \label{tab:vParams}
\centering
\begin{tabular}{l|l|l}
\hline
\multicolumn{1}{c|}{\textbf{Category}} & \multicolumn{1}{c|}{\textbf{Type}} & \multicolumn{1}{c}{\textbf{Service speed (knot)}}\\
\hline
Tanker & Crude oil carrier & 13-17\\
 & Gas tanker/LNG carrier & 16-20\\
 & Product & 13-16\\
 & Chemical & 15-18\\
\hline
Bulk carrier & Ore carrier & 14-15\\
 & Regular & 12-15\\
\hline
Container & Line carrier & 20-23\\
 & Feeder & 18-21\\
\hline
General cargo & General cargo & 14-20\\
 & Coaster & 13-16\\
\hline
Roll-on/roll-off cargo & Ro-Ro/Ro-Pax & 18-23\\
\hline
Passenger ship & Cruise ship & 20-23\\
 & Ferry & 16-23\\
\hline
\end{tabular}
\end{table}

\subsubsection{Uncertainty due to Human Error}
AIS data, excluding dynamic information, is not automatically updated by the sensors, but it is logged by the ship's crew manually, so there is a possibility of human error. This includes information such as the draft, navigation status, destination, and estimated time of arrival (ETA) of the ship. Although it is difficult to clearly distinguish the incorrectly entered information, it is possible to indirectly determine whether the manual input values have been updated using the automatically logged dynamic information. Each number in navigation status represents ship activity such as `under way using engine (0)', `at anchorage (1)', and `moored (5)'. If this field is being updated normally, it should be `0' if the ship is in-trip and `5' if it is at berth. If the navigation status of the collected AIS data is `1' or `5' above a certain GPS speed (or speed-over-ground), or if the state is set to `0' even when the speed is 0 and the location is within the port, the AIS data have not been updated on time and other manually entered information should also be questioned.

\subsection{Noon Report Data}
%  \textcolor{red}{(Will be revised in sentence form)}
Ships engaged in international navigation of more than 500 gross tons are required to send a noon report to the company, which briefly records what happened on the ship from previous noon to present noon. The noon report must basically contain sufficient information regarding the location, course, speed, and internal and external conditions affecting the vessel's voyage. Additionally, the shipping company collects information related to fuel consumption and remaining fuel onboard, propeller slip, average RPM, etc. as needed. Such information is often used as a ship's management tool and reference data, such as monitoring and evaluating ship's performance, calculating energy efficiency operating indicators, and obtaining fuel and freshwater order information. Despite its customary use, the standardized information in the noon reports may not be sufficient to accurately assess the performance of the ship, due to several problems discussed as follows. This information is based on the average values from noon to noon. For an accurate ship performance analysis, higher frequency samples and additional data may be recommended.
% In accordance with Article 28 of SOLAS V, 
\subsubsection{Uncertainties due to Averaging Measurements \& Human Error} \label{sec:noonReportsAvgProb}
Basically, information reported through the noon reports is created based on the measurement values of the onboard sensor. Therefore, it also has the possibility to involve the problem of inherently faulty sensors and incorrect measurements, as discussed in section \ref{sec:incorrMeasureInServData}. Apart from the problems caused by sensors, the noon report data may have problems caused by the use of 24-hour averaged values and human errors. The data collection interval is once a day and the average of the values recorded for 24 hours is reported, thus, significant inaccuracies may be included in the data. \citet{aldous2015uncertainty} performed a sensitivity analysis to assess the uncertainty due to the input data for ship performance analysis using continuously recorded in-service data and noon reports. It was observed here that the uncertainty of the outcome was significantly sensitive to the number of samples in the dataset. In other words, such uncertainty can be mitigated through the use of data representing longer time-series, data collection with higher frequency, and data processing. These results were also confirmed by \citet{park2017comparative} and \citet{themelis2018comparative}. \citet{park2017comparative} demonstrated in a case study that the power consumption between the noon reports and the recorded sensor data differed by 6.2\% and 17.8\% in ballast and laden voyage, respectively. 

%compared continuously recorded in-service ship data with noon reports and observed that \textcolor{red}{the number of observations in the dataset had a significant impact on uncertainty (what do u mean!?)}, and 

Using the averaged values over a long time period, as in the case of noon reports, the variations due to acceleration/deceleration and maneuvering cannot be captured. In particular, in the case of ships that sail relatively short voyages such as feeder ships and ferries, inappropriate data for performance analysis may be provided due to frequent changes in the operational state. In the case of information regarding the weather and sea states, the information generally corresponds to the condition right before the noon report is sent from the ship, therefore, it is not easy to account for the changes in the performance of the ship due to the variation of weather conditions during the last 24 hours. In general, the information to be logged in the noon report is read and noted by a person from onboard sensors. Thus, it is possible that the time at which the values are read from the sensors everyday may be different as well as different sensors may be used for the values to be logged for the same field. In addition, there may be cases when the observed value is incorrectly entered into the noon report. Thus, if the process of preparing the noon reports is not automated, there would always be a possibility of human errors in the data.

\section{Results: Data Processing Framework} \label{sec:results}

\begin{figure}
\centering

\begin{tikzpicture}[font=\small,thick, node distance = 0.35cm]

\node[draw,
    rounded rectangle,
    minimum width = 2.5cm,
    minimum height = 1cm
] (block1) {Raw Data};

\node[draw,
    below=of block1,
    minimum width=3.5cm,
    minimum height=1cm,
    align=center
] (block2) {Ensure Uniform \\ Time Steps};

\node[draw,
    below=of block2,
    minimum width=3.5cm,
    minimum height=1cm
] (block3) {Divide into Trips};

\node[draw,
    below=of block3,
    minimum width=3.5cm,
    minimum height=1cm,
    align=center
] (block4) {Interpolate Hindcast \\ (Using GPS Data)};

\node[draw,
    trapezium, 
    trapezium left angle = 65,
    trapezium right angle = 115,
    trapezium stretches,
    left=of block4,
    minimum width=3.5cm,
    minimum height=1cm
] (block5) {Weather Hindcast};

\node[draw,
    below=of block4,
    minimum width=3.5cm,
    minimum height=1cm
] (block6) {Derive New Features};

\node[draw,
    diamond,
    right=of block6,
    minimum width=2.5cm,
    inner sep=1,
    align=center
] (block17) {Interpolation \\ Error?};

\node[draw,
    below=of block6,
    minimum width=3.5cm,
    minimum height=1cm
] (block7) {Validation Checks};

\node[draw,
    diamond,
    below=of block7,
    minimum width=2.5cm,
    inner sep=1,
    align=center
] (block8) {Data Processing \\ Errors Detected?};

\node[coordinate,right=1.8cm of block8] (block9) {};
\node[coordinate,right=1.6cm of block4] (block10) {};

\node[draw,
    below=of block8,
    minimum width=3.5cm,
    minimum height=1cm
] (block11) {Fix Draft \& Trim};

\node[draw,
    below=of block11,
    minimum width=3.5cm,
    minimum height=1cm,
    align=center
] (block12) {Calculate Hydrostatics \\ (Displacement, WSA, etc.)};

\node[draw,
    trapezium, 
    trapezium left angle = 65,
    trapezium right angle = 115,
    trapezium stretches,
    left=of block12,
    minimum width=3.5cm,
    minimum height=1cm
] (block15) {Ship Particulars};

\node[draw,
    below=of block12,
    minimum width=3.5cm,
    minimum height=1cm,
    align=center
] (block13) {Calculate Resistance \\ Components};

\node[draw,
    below=of block13,
    minimum width=3.5cm,
    minimum height=1cm,
    align=center
] (block16) {Data Cleaning \& \\ Outlier Detection};

\node[draw,
    rounded rectangle,
    below=of block16,
    minimum width = 2.5cm,
    minimum height = 1cm,
    inner sep=0.25cm
] (block14) {Processed Data};

% Arrows
\draw[-latex] (block1) edge (block2)
    (block2) edge (block3)
    (block3) edge (block4)
    (block4) edge (block6)
    (block6) edge (block7)
    (block7) edge (block8)
    (block8) edge node[anchor=east,pos=0.25,inner sep=2.5]{No} (block11)
    (block11) edge (block12)
    (block12) edge (block13)
    (block13) edge (block16)
    (block16) edge (block14);

\draw[-latex] (block5) edge (block4);
\draw[-latex] (block15) edge (block12);

\draw[-latex] (block8) -| (block9) node[anchor=south,pos=0.1,inner sep=2.5]{Yes}
    (block9) -| (block17);

\draw[-latex] (block17) |- (block10) 
    (block10) |- (block4) node[anchor=south,pos=0.1,inner sep=2.5]{Yes};

\draw[-latex] (block17) -- (block6) node[anchor=south,pos=0.4,inner sep=2.5]{No};

\end{tikzpicture}
\caption{Data processing framework flow diagram.} \label{fig:flowDiag}
\end{figure}
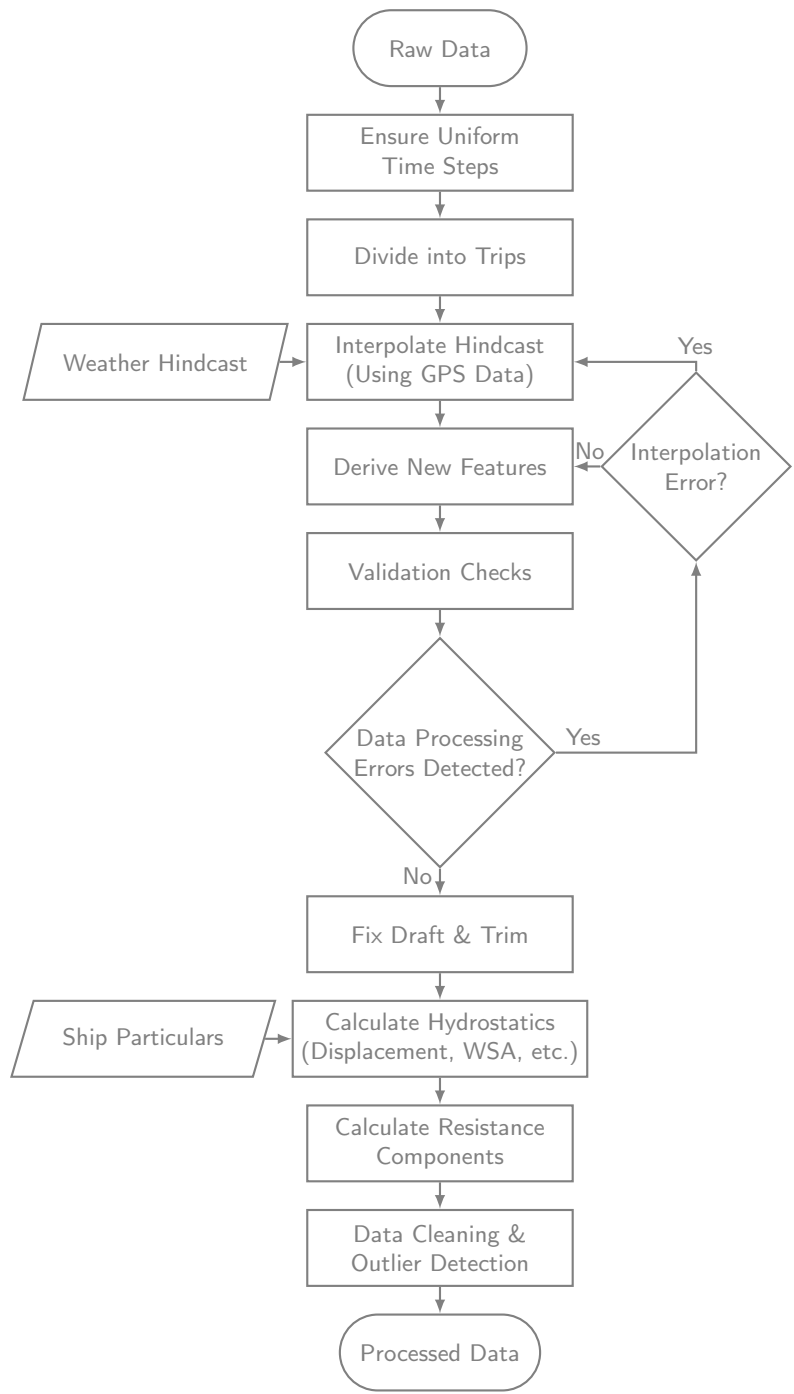

The results here are presented in the form of the developed data processing framework, which can be used to process raw data obtained from one of the above mentioned data sources (in section \ref{sec:dataSources}) for ship performance analysis. The data processing framework is designed to resolve most of the problems cited in the above section. Figure \ref{fig:flowDiag} shows the flow diagram for the data processing framework. The following sections will explain briefly the consecutive processing steps of the given flow diagram. It may be possible that the user may not able to carry-out each step due to unavailability of some information or features in the dataset, for example, due to the unavailability of the GPS data (latitude, longitude and timestamp variables), it may not be possible to interpolate weather hindcast data. In such a case, it is recommended to skip the corresponding step and continue with the next one. 

The data processing framework has been outlined in such a manner that, after being implemented, it can be executed in a semi-automatic manner, i.e., requiring limited intervention from the user. The semi-autonomous nature of the framework would also result in fast data processing, which can be important for very large datasets. The implementation of the framework in terms of executable code is also quite important to obtain a semi-automatic and fast implementation of the data processing framework. Therefore, it is recommended to adopt best practices and optimized algorithms for each individual processing step according to the programming language in use. On another note, the reliability of the data processing activity is also quite critical to obtain good results. Therefore, it is important to carry-out the validation of work done in each processing step by creating visualization (or plots) and inspecting them for any undesired errors. The usual practice adopted here, while processing the data using the framework, is to create several such visualizations, like time-series plots of data variables in trip-wise manner (explained later in section \ref{sec:divideIntoTrips}), at the end of each processing step and then inspecting them to validate the outcome. 

\subsection{Ensure Uniform Time Steps}

Ensuring uniform and evenly-spaced samples would not only make it easier to apply time-gradient-based data processing or analysis steps but would also help avoid any misunderstanding while visualizing the data, by clearly showing a gap in the time-series plots (when plotted against sample numbers) and removing any abrupt jumps in the data values. Depending on the data acquisition (DAQ) system, the in-service data recorded onboard a ship is generally recorded with a uniform and evenly spaced sampling interval. Nevertheless, it is observed that the extracted sub-dataset from the main database may contain several missing time steps (or timestamps). In such a case, it is recommended to check for such missing timestamps by simply calculating the gradient of timestamps, and for each missing timestamp, just add an empty row consisting only the missing timestamp value. Finally, the dataset should be sorted according the timestamps, resulting in a uniform and evenly-spaced list of samples. 

\begin{figure}[ht]
\centering
\includegraphics[width=0.5\linewidth]{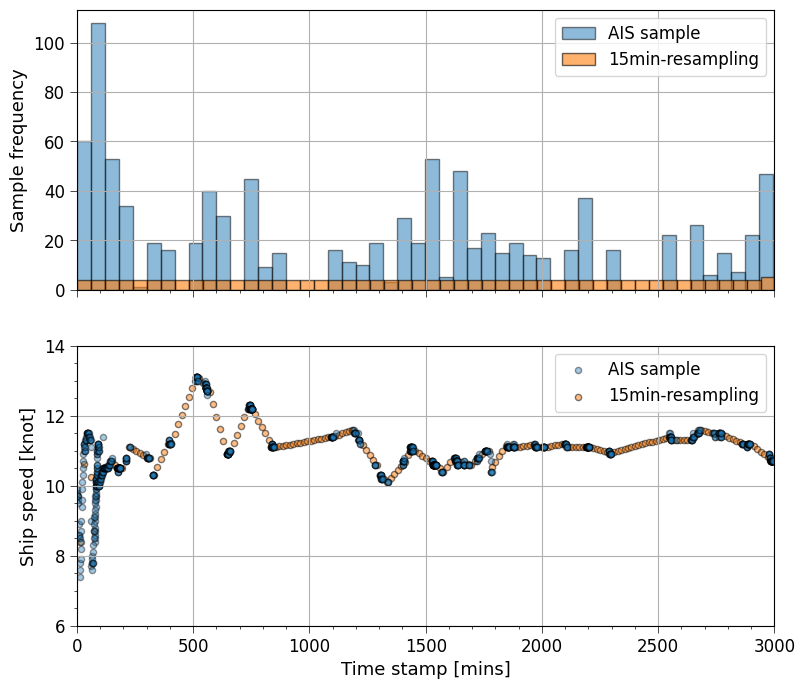}
\caption{Down-sampling the collected AIS data to 15 minutes interval.} \label{fig:resampleSOG}
\end{figure}

Similar procedure can be adopted for a noon report dataset. The noon reports are generally recorded every 24 hours, but it may sometimes be more or less than 24 hours if the vessel's local time zone is adjusted, specially on the day of arrival or departure. The same procedure may not be feasible in case of AIS data, as the samples here are sporadically distributed in general. Here, the samples are collected at different frequencies depending on the ship's moving state, surrounding environment, traffic, and the type of AIS receiving station (land-based or satellite). It is observed here that the data is collected in short and continuous sections of the time-series, leaving some large gaps between samples, as shown in figure \ref{fig:resampleSOG}. Here, it is recommended to first resample the short and continuous sections of AIS data to a uniform sampling interval through data resampling techniques, i.e., up-sampling or down-sampling (as demonstrated by \citet{virtanen2020scipy}), and then, fill the remaining large gaps with empty rows.

\subsection{Divide Into Trips} \label{sec:divideIntoTrips}

Using conventional tools, data visualization becomes a challenge if the number of samples in the dataset is enormously large. It may simply not be practical to plot the whole time-series in a single plot. Moreover, dividing the time-series into individual trips may be considered as neat and help discretize the time-series into sensible sections which may be treated individually for further data processing and analysis. Plotting an individual trip would also give a complete overview of a port-to-port journey of the ship. Dividing the data into trips and at-berth legs would also make further data processing computationally less expensive as it may be possible to ignore a large number of samples (for further steps) where the ship is not in a trip. For such samples, it may not be necessary to interpolate hindcast, calculate hydrostatics, calculate resistance components, etc. Lastly, identifying individual trips would also make draft and trim correction step easier.

\begin{figure}[ht]
    \centering
    \begin{subfigure}{0.48\linewidth}
        \includegraphics[width=\linewidth]{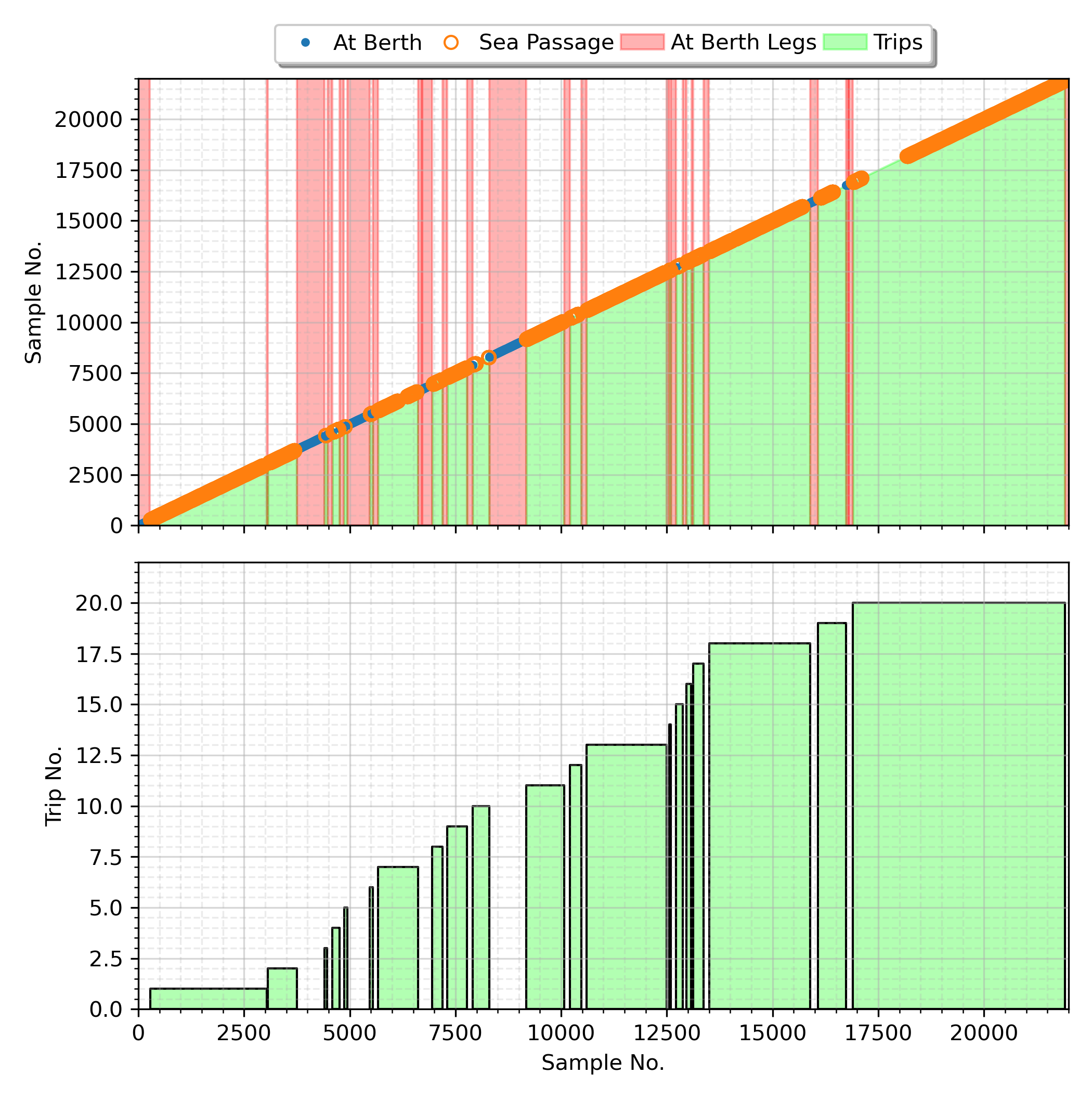}
        \caption{Splitting time-series into trips using the `State' variable.} \label{fig:splitTSviaState}
    \end{subfigure}
    \begin{subfigure}{0.48\linewidth}
        \includegraphics[width=\linewidth]{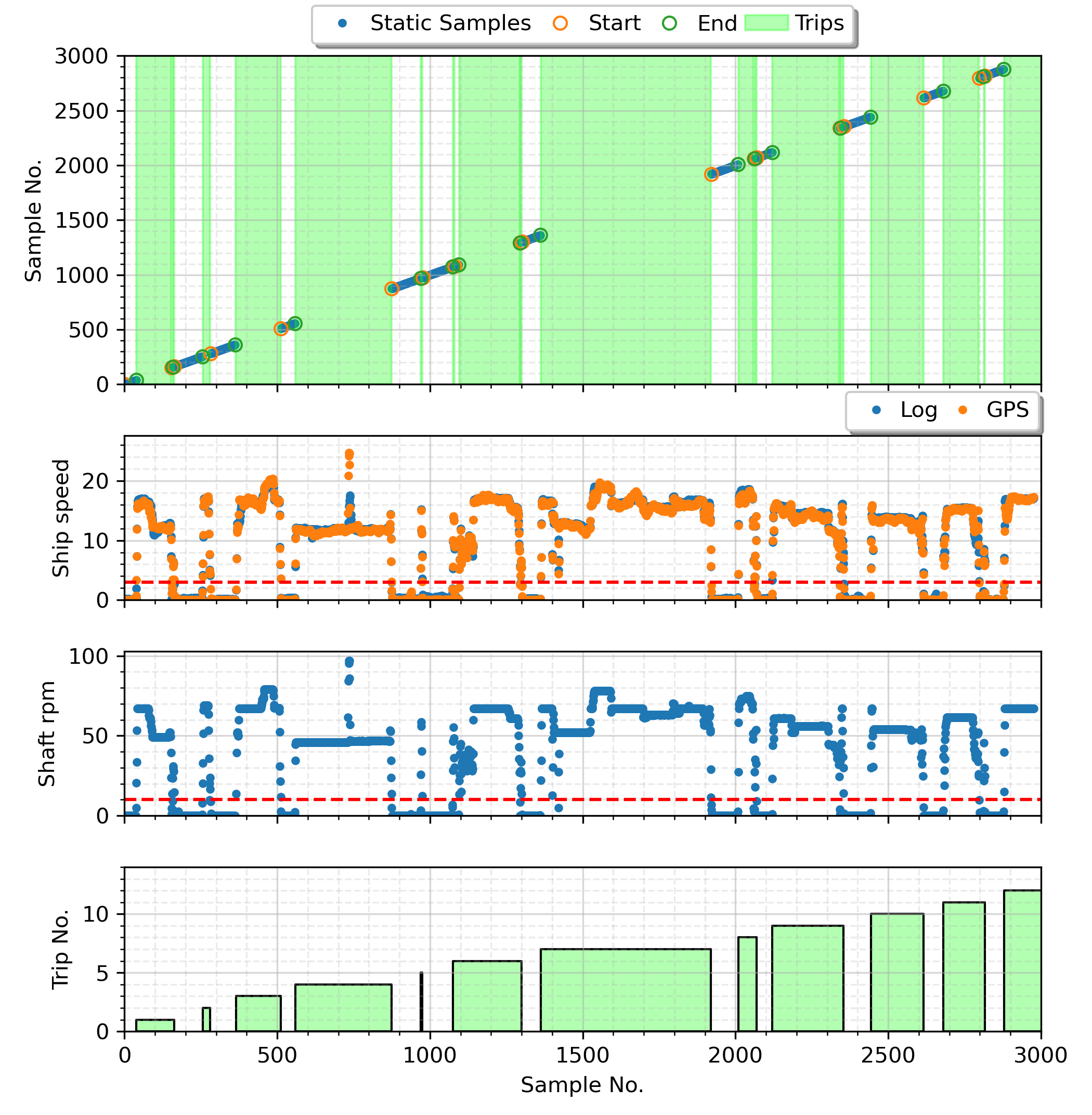}
        \caption{Splitting time-series into trips using threshold values (indicated by dashed red lines) for shaft rpm (10 rpm) and GPS speed (3 knots) variables.} \label{fig:splitTS}
    \end{subfigure}
    \caption{Splitting time-series into trips.}
\end{figure}

Dividing data into trips is substantially easier for noon reports and AIS data as they are generally supplied with a source and/or destination port name. In case of in-service data, it may be possible that no such information is available. In such a case, if the GPS data (latitude and longitudes) is available, it may be possible to just plot the samples on the world map and obtain individual trips by looking at the port calls. Alternatively, if the in-service data is supplied with a `State' variable\footnote{Generally available for ships equipped with Marorka systems (www.marorka.com).} (mentioned by \citet{Gupta2019}), indicating the propulsive state of the ship, like `Sea Passage', `At Berth', `Maneuvering', etc., it is recommended to find the continuous legs of `At Berth' state and enumerate the gaps in these legs with trip numbers, containing the rest of the states, as shown in figure \ref{fig:splitTSviaState}. Otherwise, it is recommended to use the shaft rpm and GPS speed (or speed-over-ground) time-series to identify the starting and end of each port-to-port trip. Here, a threshold value can be adopted for the shaft rpm and GPS speed. All the samples above these threshold values (either or both) are considered to be in-trip samples, as shown in figure \ref{fig:splitTS}. Thus, continuous legs of such in-trip samples can simply be identified and enumerated. It may also be possible to append few samples before and after each of these identified trips to obtain a proper trip, starting from zero and ending at zero. Such a process is designed keeping in mind the noise in the shaft rpm and GPS speed variables when the ship is actually static. Finally, if the GPS data is available, further adjustments can be done by looking at the port calls on the world map plotted with the GPS data.

\subsection{Interpolate Hindcast \& GPS Position Correction} \label{sec:interpolateHindcast}

Even if the raw data contains information regarding the state of the weather for each data sample, it may be a good idea to interpolate weather hindcast (or metocean) data available from one of the well-established sources. The interpolated hindcast data would not only provide a quantitative measure of the weather conditions (and, consequently, the environmental loads) experienced by the ship, but it would also help carry-out some important validation checks (discussed later in section \ref{sec:resultsValChecks}). In order to interpolate hindcast data, the information regarding the location (latitude and longitude) and recording timestamp must be available in the ship's dataset. For ship performance analysis, it should be aimed that, at least, the information regarding the three main environmental load factors, i.e., wind, waves and sea currents, is gathered from the weather hindcast sources. For a further detailed analysis, it may also be a good idea to obtain additional variables, like sea water temperature (both surface and gradient along the depth of the ship), salinity, etc.

\begin{figure}[ht]
    \centering
    \begin{subfigure}{0.48\linewidth}
        \includegraphics[width=\linewidth]{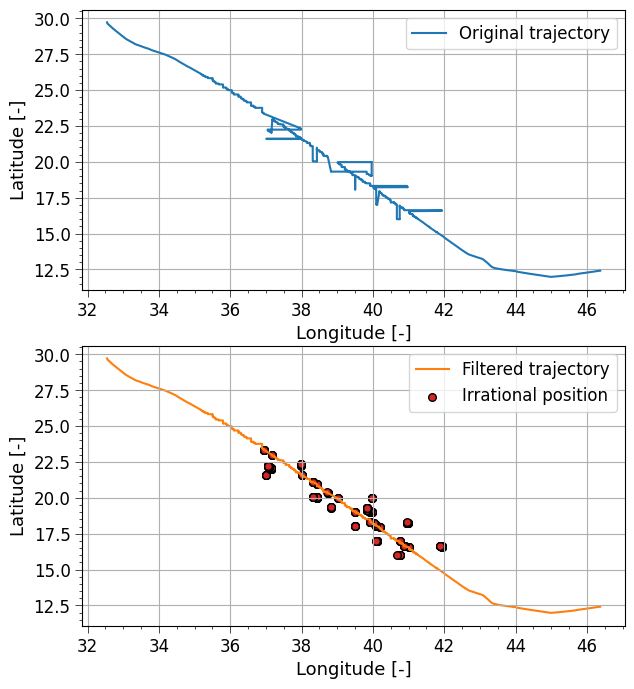}
        \caption{Original trajectory and filtered trajectory with irrational GPS position.} \label{fig:gps_outlier}
    \end{subfigure}
    \begin{subfigure}{0.48\linewidth}
        \includegraphics[width=\linewidth]{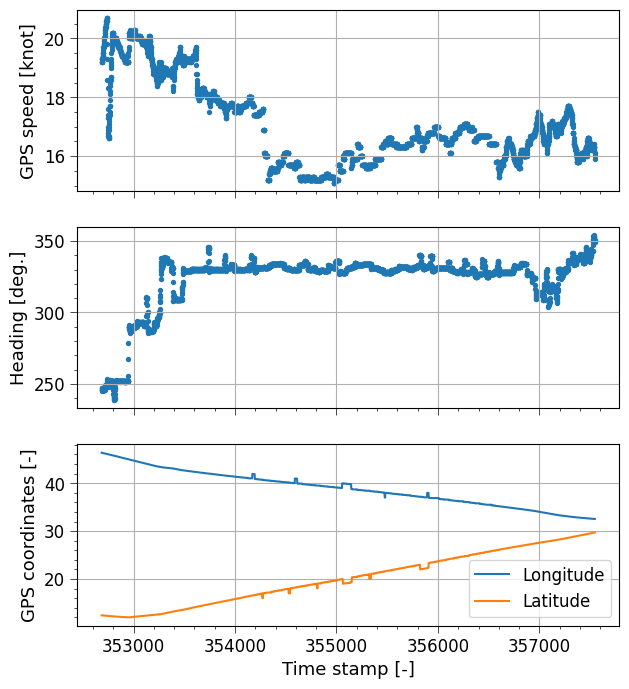}
        \caption{Trends of GPS speed, heading, and position of the ship according to the corresponding period.} \label{fig:gps_condition}
    \end{subfigure}
    \caption{GPS position cleaning using the steady-state detection algorithm.}
\end{figure}

Before interpolating the weather hindcast data to the ship's location and timestamps, it is recommended to ensure that the available GPS (or navigation) data is validated and corrected (if possible) for any errors. If the GPS data is inaccurate, weather information at the wrong location is obtained, resulting in incorrect values for further analysis. For instance, the ship's original trajectory obtained from the GPS data, presented in figure \ref{fig:gps_outlier}, shows that the ship proceeds in a certain direction while suddenly jumping to an off-route location occasionally. The ship, of course, may have gone off-route as shown here, but referring to the GPS speed and heading of the ship at the corresponding time, shown in figure \ref{fig:gps_condition}, it is obvious that the navigation data is incorrect. Here, such an irrational position change can be detected through the two-stage steady-state (or stationarity) filter suggested by \citet{Gupta2021}, based on the method developed by \citet{Dalheim2020}. The first stage of the filter uses a sliding window to remove unsteady samples by performing a t-test on the slope of the data values, while the second stage performs an additional gradient check for the samples failing in the first stage to retain the misidentified samples. The `irrational position' in figure \ref{fig:gps_outlier} shows the coordinates identified as unsteady when the above two-stage filter is applied to longitude and latitude time-series. The filtered trajectory is further obtained after removing the samples with `irrational position' from the original data. 
% This filter is found to be very effective when detecting sudden fluctuations rather than steady changes in the time-series data.

The hindcast data sources generally allow downloading a subset of the variables, timestamps, and a sub-grid of latitudes and longitudes, i.e., the geographical location. Depending on the hindcast source, the datasets can be downloaded manually (by filling a form), using an automated API script, or even by directly accessing their ftp servers. It may also be possible to select the temporal and spatial resolution of the variables being downloaded. In some cases, the hindcast web servers allows the users to send a single query, in terms of location, timestamp, and list of variables, to extract the required data for an individual sample. But every query received by these servers is generally queued for processing, causing substantially long waiting times, as they are facing a good amount of traffic from all over the world. Thus, it is recommended to simply download the required subset of data on a local machine for faster interpolation.   

Once the hindcast data files are available offline, the main task at hand is to understand the cryptic (but highly efficient) data packaging format. Now-a-days, the two most poplar formats for such data files are GRIdded Binary data (GRIB) and NetCDF. GRIB (available as GRIB1 or GRIB2) is the international standard accepted by World Meteorological Organization (WMO), but due to some compatibility issues with windows operating systems, it may be preferable to use the NetCDF format.

% The data format can be understood by reading the `dimensions' and `variables' information available in the downloaded file, i.e., `wData.dimensions' and `wData.variables', respectively. Here, `wData' is the name of the variable with which the weather data file is loaded. Each of these data files would contain the mapping variables (or dimensions), like latitude, longitude, time, depth, etc., and the data variables, like 10 metre U wind component (u10), 10 metre V wind component (v10), etc. To read the value of a data variable in a file, the indices (or range of indices) of the mapping variables have to be supplied in a manner similar to accessing values in a multidimensional array. For example, to extract the values of `u10' over a grid of latitudes and longitudes for a particular timestamp, the user needs to supply the index of the timestamp of interest in the given mapping (say, 0, i.e., the first timestamp in the file) and the range of indices of latitude (say, 100 to 200) and longitude (say, 200 to 300), which can be queried in python as wData[`u10'][0, 100:200, 200:300], provided the variable `u10' is stored in (`time', `latitude', `longitude') format with respect to the mapping variables. Similarly, to extract the whole grid (available in the data file) for the first timestamp, the range of latitude and longitude can be just represented using the colon (:) operator in python, i.e., wData[`u10'][0, :, :].

Finally, a step-by-step interpolation has to be carried-out for each data sample from the ship's dataset. Algorithm \ref{algo:hindcastInterp} shows a simple procedure for n-th order (in time) interpolation scheme. Here, the spatial and temporal interpolation is performed in steps \ref{algoStep:spatialInterp} and \ref{algoStep:temporalInterp}, respectively. For a simple and reliable procedure, it is recommended to perform the spatial interpolation using a grid of latitudes and longitudes around the ship's location, after fitting a linear or non-linear 2D surface over the hindcast grid. It may be best to use a linear surface here as, firstly, the hindcast data may not be so accurate that performing a higher order interpolation would provide any better estimates, and secondly, in some case, higher order interpolation may result in highly inaccurate estimates, due to the waviness of the over-fitted non-linear surface. Similar arguments can be given in the case of temporal interpolation, and therefore, a linear interpolation in time can also be considered acceptable. The advantage of using the given algorithm is that the interpolation steps, here, can be easily validated by plotting contours (for spatial interpolation) and time-series (for temporal interpolation). 

\begin{algorithm}
\caption{A simple algorithm for n-th order interpolation of weather hindcast data variables.}\label{algo:hindcastInterp}
\begin{algorithmic}[1]

\State $wData \gets $ weather hindcast data
\State $x \gets $ data variables to interpolate from hindcast
\State $wT \gets $ timestamps in $wData$

\ForAll{timestamps in ship's dataset}
    
    \State $t \gets $ current ship time stamp
    \State $loc \gets $ current ship location (latitude \& longitude)
    \State $i \gets n+1$ indices of $wT$ around $t$
    
    \ForAll{$x$}
        \ForAll{$i$}
            \State $x[i] \gets $ 2D spatial interpolation at $loc$ using $wData[x][i, :, :]$ \label{algoStep:spatialInterp}
        \EndFor
        \State $X \gets $ n-th order temporal interpolation at $t$ using $x[i]$ \label{algoStep:temporalInterp}
    \EndFor

\EndFor
\end{algorithmic}
\end{algorithm}

An important feature of hindcast datasets is masking the invalid values. For instance, the significant wave height should only be predicted by the hindcast model for the grid nodes which fall in the sea, requesting the value of such a variable on land should result in an invalid value. Such invalid values (or nodes) are by default masked in the downloaded hindcast data files, probably for an efficient storage of the data. These masked nodes may be filled with zeros before carrying-out the spatial interpolation in step \ref{algoStep:spatialInterp}, as one or more of these nodes may be contributing to the interpolation. Alternatively, if a particular masked node is contributing to the interpolation, it can be set to the mean of other nodes surrounding the point of interpolation, as suggested by \citet{Ejdfors2019}. It is argued by \citet{Ejdfors2019} that this would help avoid the artificially low (zero) values during the interpolation, but if the grid resolution is fine-enough, it is expected that the calculated mean (of unmasked surrounding nodes) would also not be much higher than zero. 

\subsection{Derive New Features}

Interpolating the weather hindcast variables to ship's location at a given time would provide the hindcast variables in the global (or the hindcast model's) reference frame. For further analysis, it may be appropriate to translate these variables to ship's frame of reference, and furthermore, it may be desired to calculate some new variables which could be more relevant for the analysis or could help validate the assimilated (ship and hindcast) dataset. The wind and sea current variables, obtained from the hindcast source and the ship's dataset, can be resolved into the longitudinal and transverse speed components for validation and further analysis. Unfortunately, the wave load variables cannot be resolved in a similar manner, but the mean wave direction should be translated into the relative mean wave direction (relative to the ship's heading or course).   

\subsection{Validation Checks} \label{sec:resultsValChecks}

Although it is recommended to validate each processing step by visualizing (or plotting) the task being done, it may be a good idea to take an intermediate pause and perform all types of possible validation checks. These validation checks would not only help assess the dataset from reliability point of view but can also be used to understand the correlation between various features. The validation checks can be done top-down, starting from the most critical feature to the least one. As explained in section \ref{sec:bestPractices}, the shaft power measurements can be validated against the shaft rpm and shaft torque measurements, if these are available, else just plotting the shaft rpm against the shaft power can also provide a good insight into the quality of data. For a better assessment, it is suggested to visualize the shaft rpm vs shaft power overlaid with the engine operational envelope and propeller curves, as presented by \citet{Liu2020} (in figure 11). Any sample falling outside the shaft power overload envelope (specially at high shaft rpm) should be removed from the analysis, as they may be having measurement errors. It may also be possible to make corrections, if the shaft power data seems to be shifted (up or down) with respect to the propeller curves due to sensor bias.

\begin{figure}[ht]
\centering
\includegraphics[width=0.6\linewidth]{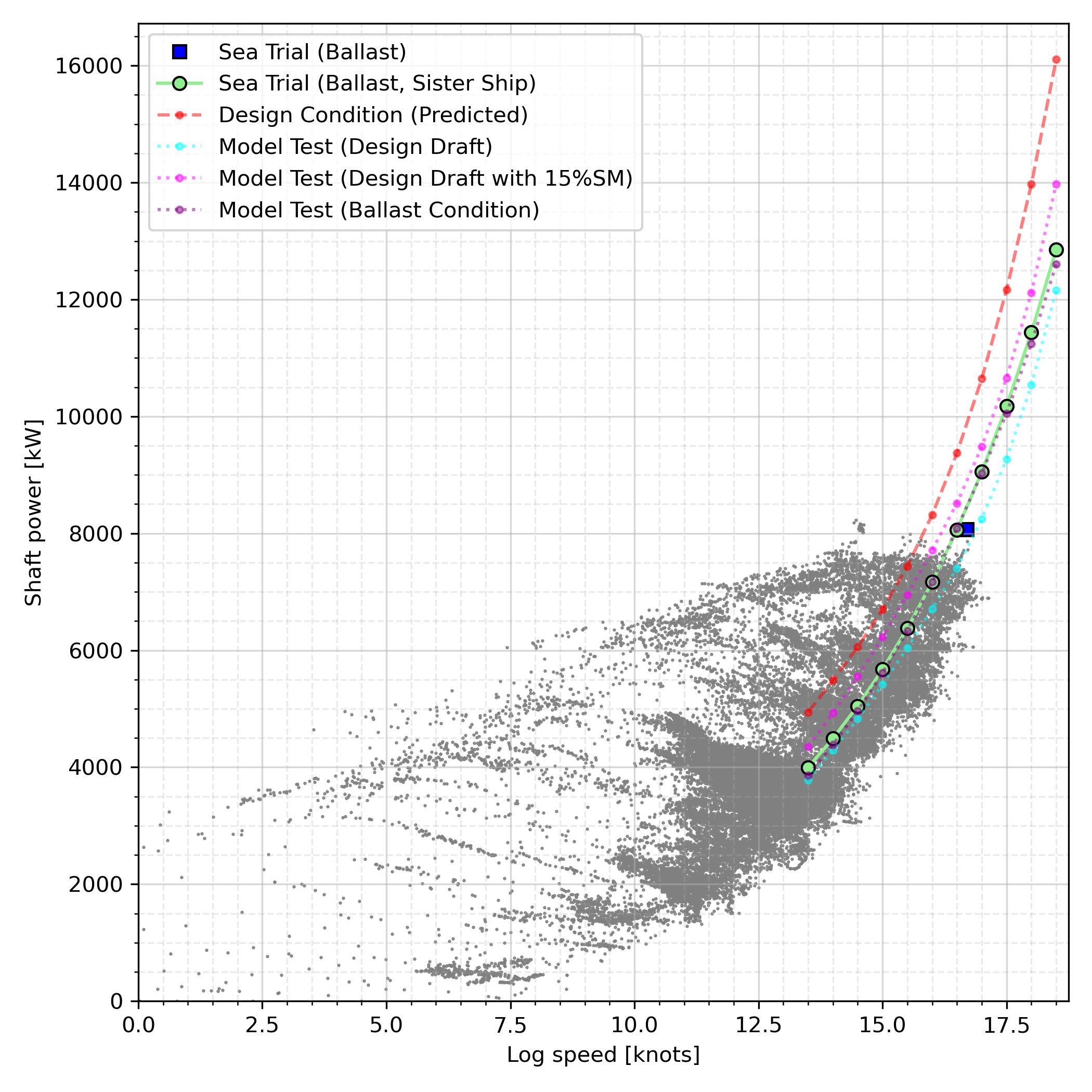}
\caption{Speed-though-water (log speed) vs shaft power with various estimates of speed-power calm-water curves.} \label{fig:speedVsPowerWSPCurves}
\end{figure}

The quality of speed-through-water measurements can be assessed by validating it against its estimate, obtained as a difference between the speed-over-ground and longitudinal current speed. Here, it should be kept in mind that the two values may not be a very good match due to several problems cited in section \ref{sec:incorrMeasureInServData}. Visualizing the speed-though-water vs shaft power along with all the available estimates of the speed-power calm-water curve is also an important validation step (shown in figure \ref{fig:speedVsPowerWSPCurves}). Here, the majority of measurement data should accumulate around these curves. In case of disparity between the curves, the curve obtained through the sea trial of the actual ship may take precedence. 

The interpolated weather hindcast data variables must also be validated against the measurements taken onboard the ship. This is quite critical as the sign and direction notations assumed by the hindcast models and ship's sensors (or data acquisition system) are probably not the same, which may cause mistakes during the interpolation step. Moreover, most ships are generally equipped with anemometers that can measure the actual and relative wind speed and directions, and these two modes (actual or relative) can be switched through a simple manipulation by the crew onboard. It is possible that this mode change may have occurred during the data recording duration, resulting in errors in the recorded data. In addition, there may be a difference between the reference height of the wind hindcast data and the vertical position of the installed anemometer, which may lead to somewhat different results even at the same location at sea. The wind speed at the reference height (${V_{WT}}_{ref}$) can be corrected using the anemometer recorded wind speed ($V_{WT}$), assuming a wind speed profile, as follows (recommended by \citet{ITTC2017}):

\begin{equation}\label{eq:referenceHeight}
{V_{WT}}_{ref} = V_{WT}\left(\frac{Z_{ref}}{Z_{a}}\right)^{\frac{1}{9}}
\end{equation}

Where $Z_{ref}$ is the reference height above the sea level and $Z_a$ is the height of the anemometer.

\begin{figure}[ht]
\centering
\includegraphics[width=0.5\linewidth]{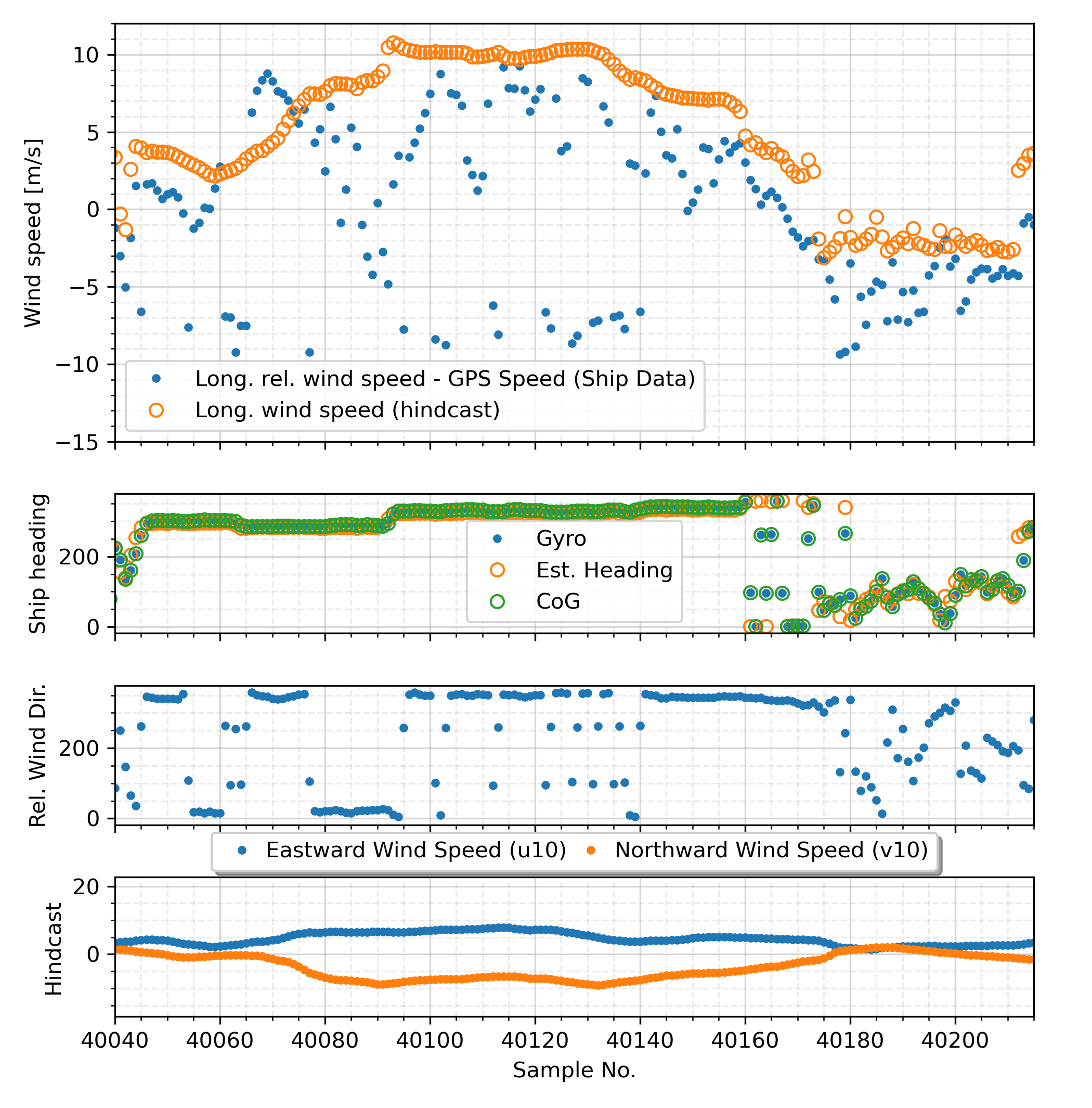}
\caption{Validating longitudinal wind speed obtained using the ship data against the values obtained from the hindcast. The time-averaging problem with angular measurements around 0 or 360 degrees (explained in section \ref{sec:timeAvgProb}) is clearly visible here.} \label{fig:longWindSpeedValidation}
\end{figure}

Finally, these wind measurements can be translated into the longitudinal and transverse relative components. The obtained transverse relative wind speed can be validated against the transverse wind speed, obtained from the hindcast source, as they are basically the same. Similarly, the difference between the longitudinal relative wind speed and the speed-over-ground of the ship can be validated against the longitudinal wind speed measurements from hindcast, as shown in figure \ref{fig:longWindSpeedValidation}. In case of time-averaged in-service data, the problem of faulty averaging of angular measurements when the measurement values are near 0 or 360 degrees (i.e., the angular limits), explained in section \ref{sec:timeAvgProb}, must also be verified and appropriate corrective measures should be taken. From figure \ref{fig:longWindSpeedValidation}, it can be clearly seen that the time-averaging problem (in relative wind direction) causes the longitudinal wind speed (estimated using the ship data) to jump from positive to negative, resulting in a mismatch with the corresponding hindcast values. In such a case, it is recommended to either fix these faulty measurements, which may be difficult as there is no proven way to do it, or just use the hindcast measurements for further analysis. 

As discussed in the case of noon reports in section \ref{sec:noonReportsAvgProb}, weather information generally refers to the state of the weather at the time when the report is logged, which is probably not the average state from noon to noon. Furthermore, the wind loads here are observed based on the Beaufort scale, therefore, the deviation may be somewhat large when converted to the velocity scale. In this case, it is recommended to consider the daily average values obtained from the weather hindcast data, over the travel region, rather than the noon report values.

\subsection{Data Processing Errors}

The validation step is very critical in finding out any processing mistakes or inherent problems with the dataset, as demonstrated in the previous section. Such problems or mistakes, if detected, must be corrected or amended for before moving forward with the processing and analysis. The main mistakes found at this step are generally either interpolation mistakes or incorrect formulation of the newly derived feature. These mistakes should be rectified accordingly, as shown in the flow diagram (figure \ref{fig:flowDiag}).  

\subsection{Fix Draft \& Trim} \label{sec:fixDraft}

\begin{figure}[ht]
    \centering
    \begin{subfigure}{0.48\linewidth}
        \includegraphics[width=\linewidth]{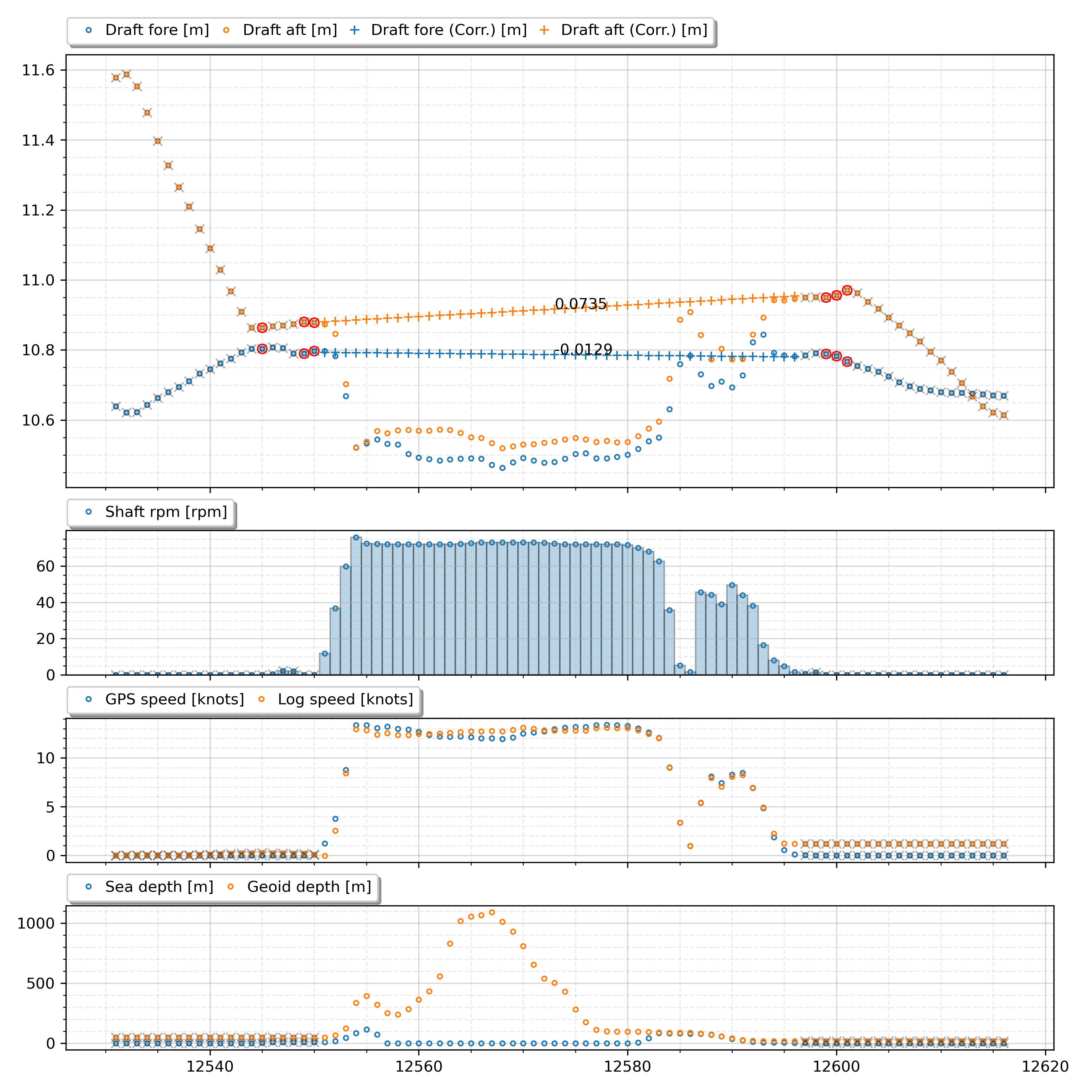}
        \caption{Simple draft correction.} \label{fig:simpleDraftCorr}
    \end{subfigure}
    \begin{subfigure}{0.48\linewidth}
        \includegraphics[width=\linewidth]{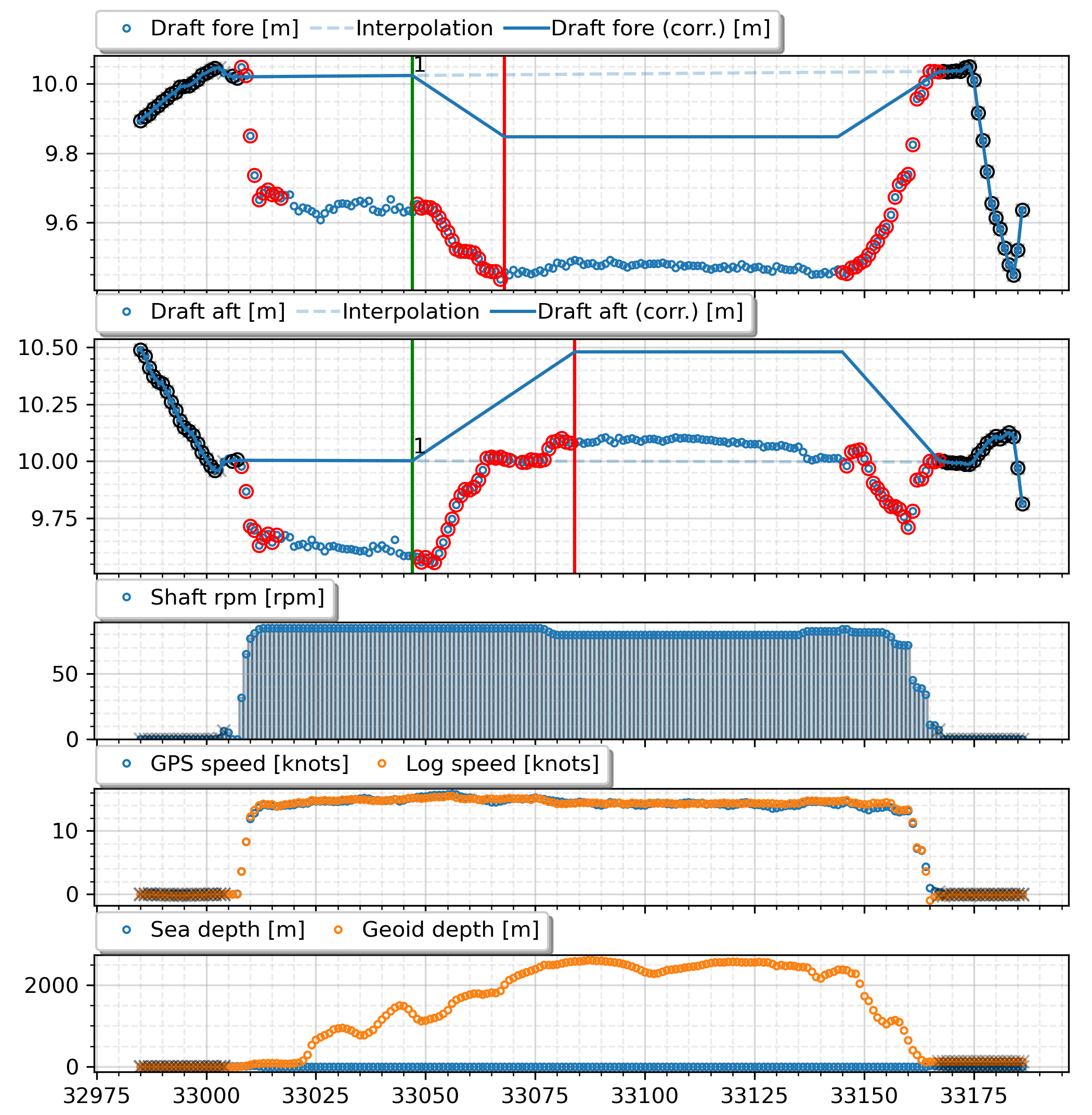}
        \caption{Complex draft correction.} \label{fig:complexDraftCorr}
    \end{subfigure}
    \caption{Correcting in-service measured draft.}
\end{figure}

The draft measurements recorded onboard the ship are often found to be incorrect due to the Venturi effect, explained briefly in section \ref{sec:incorrMeasureInServData}. The Venturi effect causes the draft measurements to drop to a lower value due to a non-zero negative dynamic pressure as soon as the ship develops a relative velocity with respect of the water around the hull. Thus, the simplest solution to fix these incorrect measurements is by interpolating the draft during a voyage using the draft measured just before and after the voyage. Such a simple solution provides good results for a simple case where the draft of the ship basically remains unchanged during the voyage, except for the reduction of draft due to consumed fuel, as shown in the figure \ref{fig:simpleDraftCorr}.

In a more complex case where the draft of the ship is changed in the middle of the voyage and the ship is still moving, i.e., conducting ballasting operations or trim adjustments during transit, the simple draft interpolation would result in corrections which can be way off the actual draft of the vessel. As shown in figure \ref{fig:complexDraftCorr}, the fore draft is seen to be dropping and the aft draft increasing in the middle of the voyage without much change in the vessel speed, indicating trim adjustments during transit. In this case, a more complex correction is applied after taking into account the change in draft during the transit. Here, first of all, a draft change operation is identified (marked by green and red vertical lines in figure \ref{fig:complexDraftCorr}), then the difference between the measurements before and after the operation is calculated by taking an average over a number of samples. Finally, a ramp is created between the starting of the draft change operation (green line) and the end of operation (red line). The slope of the ramp is calculated using the difference between the draft measurements before and after the draft change operation. The draft change operation can either be identified manually, by looking at the time-series plots, or using the steady-state (or stationarity) filter developed by \citet{Dalheim2020}.

In case of AIS data, \citet{bailey2008training} reported that 31\% of the draft information out of the investigated AIS messages had obvious errors. The draft information from AIS data generally corresponds to the condition of ships while arriving at or departing from the port, and changes due to fuel consumption and ballast adjustment onboard are rarely updated. Since the draft obtained from the AIS as well as noon reports has a long update cycle and is acquired by humans, it is practically difficult to precisely fix the draft values as in the case of in-service data. However, by comparing the obtained draft with a reference value, it may be possible to gauge whether the obtained draft is, in fact, correct. If the obtained draft excessively deviates from the reference, it may be possible to remove the corresponding data samples from further analysis or replace the obtained draft value with a more appropriate value. Table \ref{tab:draftRatio} shows the results of investigating the average draft ratio, which is the ratio of the actual draft ($T_c$) and design draft ($T_d$), for various ship types from 2013 to 2015 by \citet{olmer2017greenhouse}. As summarized in the table, the draft ratio varies depending on the ship type and the voyage type. By using these values as the above mentioned reference, the draft obtained from the AIS data and noon reports can be roughly checked and corrected.

\begin{table}[ht]
\caption{Average draft ratio ($T_c/T_d$) for different ship types. $T_c$ = actual draft during a voyage; $T_d$ = design draft of the ship.} \label{tab:draftRatio}
\centering
\begin{tabular}{l|c|c}
\hline
\multicolumn{1}{c|}{\textbf{Ship types}} & \multicolumn{1}{c|}{\textbf{Ballast Voyage}} & \multicolumn{1}{c}{\textbf{Laden Voyage}}\\
\hline
Liquefied gas tanker & 0.67 & 0.89\\
Chemical tanker & 0.66 & 0.88\\
Oil tanker & 0.60 & 0.89\\
Bulk carrier & 0.58 & 0.91\\
General cargo & 0.65 & 0.89\\
\hline
\multicolumn{3}{c}{\textit{The following ship types do not generally have ballast-only voyages.}} \\
\hline
Container  & \multicolumn{2}{c}{0.82}\\
Ro-Ro & \multicolumn{2}{c}{0.87}\\
Cruise  & \multicolumn{2}{c}{0.98}\\
Ferry pax & \multicolumn{2}{c}{0.90}\\
Ferry ro-pax & \multicolumn{2}{c}{0.93}\\
\hline
\end{tabular}
\end{table}

\subsection{Calculate Hydrostatics}

Depending on the type of performance analysis, it may be necessary to have features like displacement, wetted surface area (WSA), etc. in the dataset, as they are more relevant from a hydrodynamic point of view. Moreover, most of the empirical or physics-based methods for resistance calculations (to be done in the next step) requires these features. Unfortunately, these feature cannot be directly recorded onboard the ship. But it is fairly convenient to estimate them using the ship's hydrostatic table or hull form (or offset table) for the corresponding mean draft and trim for each data sample. Here, it is recommended to use the corrected draft and trim values, obtained in the previous step. If the detailed hull form is not available, the wetted surface area can also be estimated using the empirical formulas shown in table \ref{tab:wsaParams}. The displacement at design draft, on the other hand, can be estimated using the ship particulars and typical range of block coefficient ($C_B$), presented in table \ref{tab:cbParams}.

\begin{table}[ht]
\caption{Estimation formulas for wetted surface area of different ship types.} \label{tab:wsaParams}
\centering
\begin{tabular}{l|l|l}
\hline
\multicolumn{1}{c|}{\textbf{Category}} & \multicolumn{1}{c|}{\textbf{Formula}} & \multicolumn{1}{c}{\textbf{Reference}}\\
\hline
Tanker/Bulk carrier & $WSA = 0.99\cdot(\frac{\nabla}{T}+1.9\cdot L_{WL}\cdot T)$ & \citet{Kristensen2017} \\
Container & $WSA = 0.995\cdot(\frac{\nabla}{T}+1.9\cdot L_{WL}\cdot T)$ & \citet{Kristensen2017} \\
Other (General) & $WSA = 1.025\cdot(\frac{\nabla}{T}+1.7\cdot L_{PP}\cdot T)$ & \citet{molland2011maritime} \\
\hline
\end{tabular}
\end{table}

\begin{table}[ht]
\caption{Typical block coefficient ($C_B$) range at design draft for different ship types, given by \citet{solutions2018basic}.} \label{tab:cbParams}
\centering
\begin{tabular}{l|l|c}
\hline
\multicolumn{1}{c|}{\textbf{Category}} & \multicolumn{1}{c|}{\textbf{Type}} & \multicolumn{1}{c}{\textbf{Block coefficient ($C_B$)}}\\
\hline
Tanker & Crude oil carrier & 0.78-0.83\\
 & Gas tanker/LNG carrier & 0.65-0.75\\
 & Product & 0.75-0.80\\
 & Chemical & 0.70-0.78\\
\hline
Bulk carrier & Ore carrier & 0.80-0.85\\
 & Regular & 0.75-0.85\\
\hline
Container & Line carrier & 0.62-0.72\\
 & Feeder & 0.60-0.70\\
\hline
General cargo & General cargo/Coaster & 0.70-0.85\\
\hline
Roll-on/roll-off cargo & Ro-Ro cargo & 0.55-0.70\\
 & Ro-pax & 0.50-0.70\\
\hline
Passenger ship & Cruise ship & 0.60-0.70\\
 & Ferry & 0.50-0.70\\
\hline
\end{tabular}
\end{table}

\subsection{Calculate Resistance Components}

There are several components of the ship's total resistance, and there are several methods to estimate each of these components. The three main resistance components which generally constitutes the majority of the ship's total resistance are calm-water, added wind, and added wave resistance. It is possible to further divide the calm-water resistance into sub-components, namely, skin friction and residual resistance. The total calm-water resistance can be calculated using one of the many well-known empirical methods, like Guldhammer and Harvald (\citet{Guldhammer1970}), updated Guldhammer and Harvald (\citet{Kristensen2017}), Hollenbach (\citet{Hollenbach1998}), Holtrop and Mennen (\citet{Holtrop1982}), etc. These empirical methods are developed using the data from numerous model test results of different types of ships, and each one is proven to be fitting well on several different ship types. The latter makes choosing the right method for a ship quite complicated. 

The easiest way to choose the right calm-water resistance estimation method is to calculate the calm-water resistance from each of these methods and comparing it with the corresponding data obtained for the given ship. The calm-water data for a given ship can be obtained from the model tests, sea trial, or even filtering the operational data, obtained from one of the sources discussed here (in section \ref{sec:dataSources}), for near-calm-water condition. The usual practice here is to use the sea trial data as it is obtained and corrected for near-calm-water condition and do not suffer from scale effects, seen in model test results. But the sea trials are sometimes conducted at only the high range of speed and ballast displacement (as shown in figure \ref{fig:speedVsPowerWSPCurves}). Thus, it is recommended to use the near-calm-water filtered (and corrected) operational data to choose the right method so that a good fit can be ensured for a complete range to speed and displacement.

According to \citet{ITTC2017}, the increase in resistance due to wind loads can be obtained by applying one of the three suggested methods, namely, wind tunnel model tests, STA-JIP, and Fujiwara's method. If the wind tunnel model test results for the vessel are available, it may be considered as the most accurate method for estimating added wind resistance. Otherwise, the database of wind resistance coefficients established by STA-JIP (\citet{van2013new}) or the regression formula presented by \citet{Fujiwara2005} is recommended. From the STA-JIP database, experimental values according to the specific ship type can be obtained, whereas Fujiwara's method is based-on the regression analysis of data obtained from several wind tunnel model tests for different ship types. 

The two main sets of parameters required to estimate the added wind resistance using any of the above three methods are incident wind parameters and information regarding the exposed area to the wind. The incident wind parameters, i.e., relative wind speed and direction, can be obtained from onboard measurements or weather hindcast data. In case of weather hindcast data, the relative wind measurements can be calculated from the hindcast values according to the formulation outlined by \citet{ITTC2017} in section E.1, and in case of onboard measurements, the relative wind measurements should be corrected for the vertical position of the anemometer according to the instructions given by \citet{ITTC2017} in section E.2, also explained here in section \ref{sec:resultsValChecks}. The information regarding the exposed area to the wind can be either estimated using the general arrangement drawing of the ship or approximately obtained using a regression formula based-on the data from several ship, presented by \citet{kitamura2017estimation}.

The added wave resistance ($R_{AW}$) can also be obtained in a similar manner using one of the several well-established methods for estimating $R_{AW}$. \citet{ITTC2017} recommends conducting sea keeping model tests in regular waves to obtain $R_{AW}$ transfer functions, which can further be used to estimate $R_{AW}$ for the ship in irregular seas. To empirically obtain these transfer functions or $R_{AW}$ for a given ship, it is possible to use physics-based empirical methods like STAWAVE1 and STAWAVE2 (recommended by \citet{ITTC2017}). STAWAVE1 is a simplified method for directly estimating $R_{AW}$ in head wave conditions only, and it requires limited input, including ship's waterline length, breadth, and significant wave height. STAWAVE2 is an advanced method to empirically estimate parametric $R_{AW}$ transfer functions for a ship. The method is developed using an extensive database of sea keeping model test results from numerous ships, but unfortunately, it only provides transfer functions for approximate head wave conditions (0 to $\pm$45 degrees from bow). A method proposed by DTU (\citet{Martinsen2016}; \citet{Taskar2019}; \citet{Taskar2021}) provides transfer functions for head to beam seas, i.e., 0 to $\pm$90 degrees from bow. Finally, for all wave heading, it may be recommended to use the newly established method by \citet{Liu2020}. There have been several studies to assess and compare the efficacy of each of these methods and several other methods, but no consistent guidelines are provided regarding their applicability. 

% \begin{itemize}
%     \item Calm water resistance
%     %% Use in-service data rather than sea-trial report if it is possible.
%     %% How to determine which method is good? The accuracy of the method may depending on the ship. Try to evaluate the RMSE over the different speed sections (Add Figure)
%     \item Added wave resistance
%     %% Use added wave resistance estimation method which is available for arbitrary waves rather than the method only for head waves -> Show the differences (Add Figure)
%     \item Added wind resistance
%     %% Correct wind speed when using the onboard measurement (Add Figure: Speed profile according to the height above the water) 
%     %% If you do not know detailed hull shape above the deck -> Regression Equation from Fujiwara (\cite{kitamura2017estimation})
% \end{itemize}

\subsection{Data Cleaning \& Outlier Detection}

It may be argued by some that the process of data cleaning and outlier detection should be carried-out way earlier in the data processing framework, as proposed by \citet{Dalheim2020DataPrep}, but it should be noted here that all the above steps proposed here have to be performed only once for a given dataset, whereas data cleaning is done based on the features selected for further analysis. Since the same dataset can be used for several different analyses, which may be using different sets of features, some part of data cleaning has to be repeated before each analysis to obtain a clean dataset with as many data samples as possible. Moreover, the additional features acquired during the above listed processing steps may be helpful in determining to a better extent if a suspected sample is actually an outlier or not. 

Nevertheless, it may be possible to reduce the work load for the above processing steps by performing some basic data cleaning before some of these steps. For instance, while calculating the resistance components for in-trip data samples, it is possible to filter-out samples with invalid values for one or more of the ship data variables used to calculate these components, like speed-though-water, mean draft (or displacement), etc. This would reduce the number of samples for which the new feature has to be calculated. It should also be noted that even if such simple data cleaning (before each step) is not performed, these invalid samples would be easily filtered-out in the present step. Thus, the reliability and efficacy of the data processing framework is not affected by performing the data cleaning and outlier detection step at the end.

Most of the methods developed for ship performance monitoring assumes that the ship is in a quasi-steady state for each data sample. The quasi-steady assumption indicates that the propulsive state of the ship remains more or less constant during the sample recording duration, i.e., the ship is neither accelerating nor decelerating. This is specially critical for aforementioned time-averaged datasets, as the averaging duration can be substantially long, hiding the effects of accelerations and decelerations. Here, the two-stage steady-state filter, explained in section \ref{sec:interpolateHindcast}, can be applied to the shaft rpm time-series to remove the samples with accelerations and decelerations, resulting in quasi-steady samples. In tandem to the steady-state filter on the shaft rpm time-series, it may also be possible to use the steady-state filter, with relaxed setting, on the speed-over-ground time-series to filter-out the sample where the GPS speed (or speed-over-ground) signal suddenly drops or recovers from a dead state, resulting in measurement errors.

\begin{figure}[ht]
\centering
\begin{subfigure}[]{0.42\linewidth}
\includegraphics[width=\linewidth]{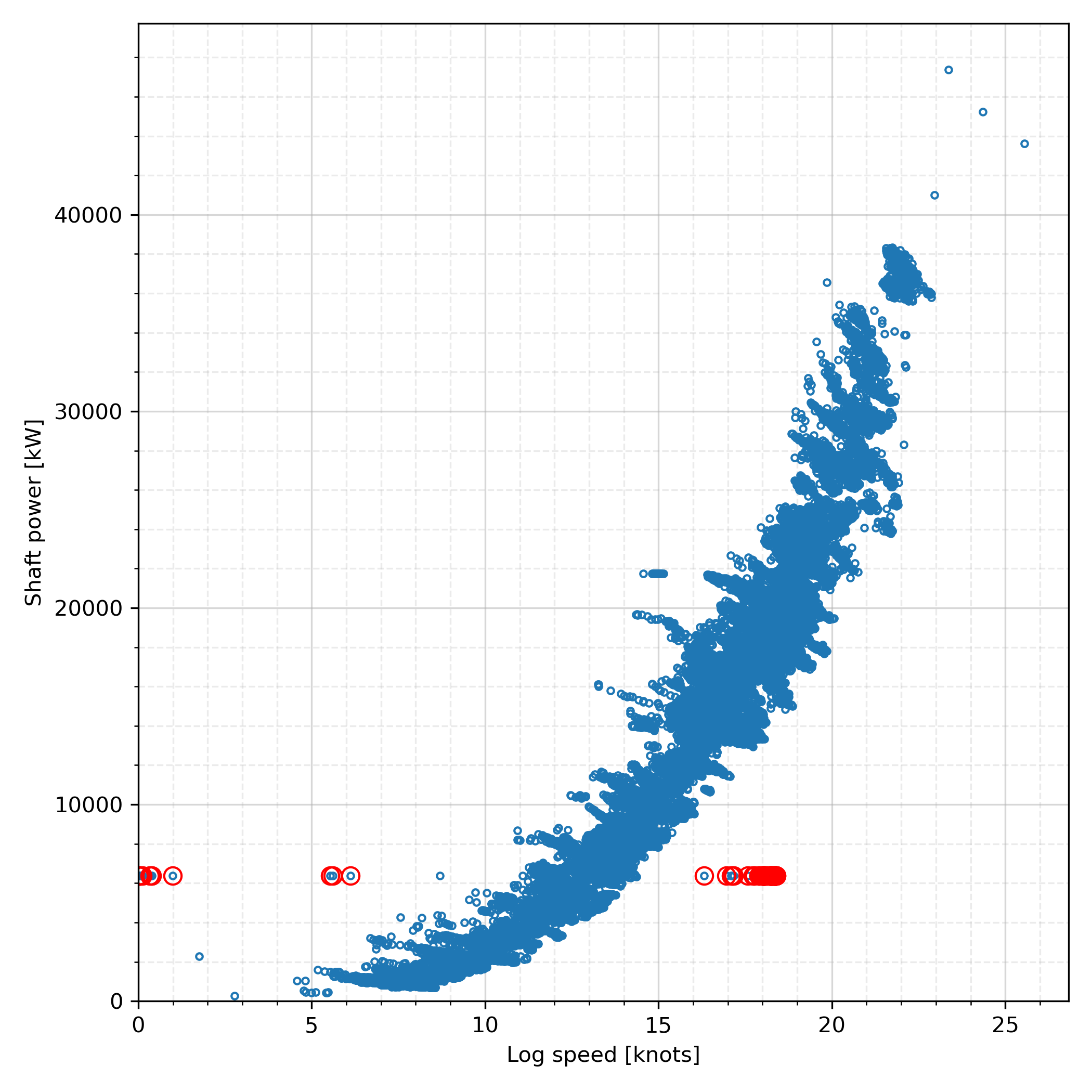}
\caption{Log speed (or stw) vs shaft power.} \label{fig:corrDefyingOutliersSP}
\end{subfigure}
\begin{subfigure}[]{0.57\linewidth}
\includegraphics[width=\linewidth]{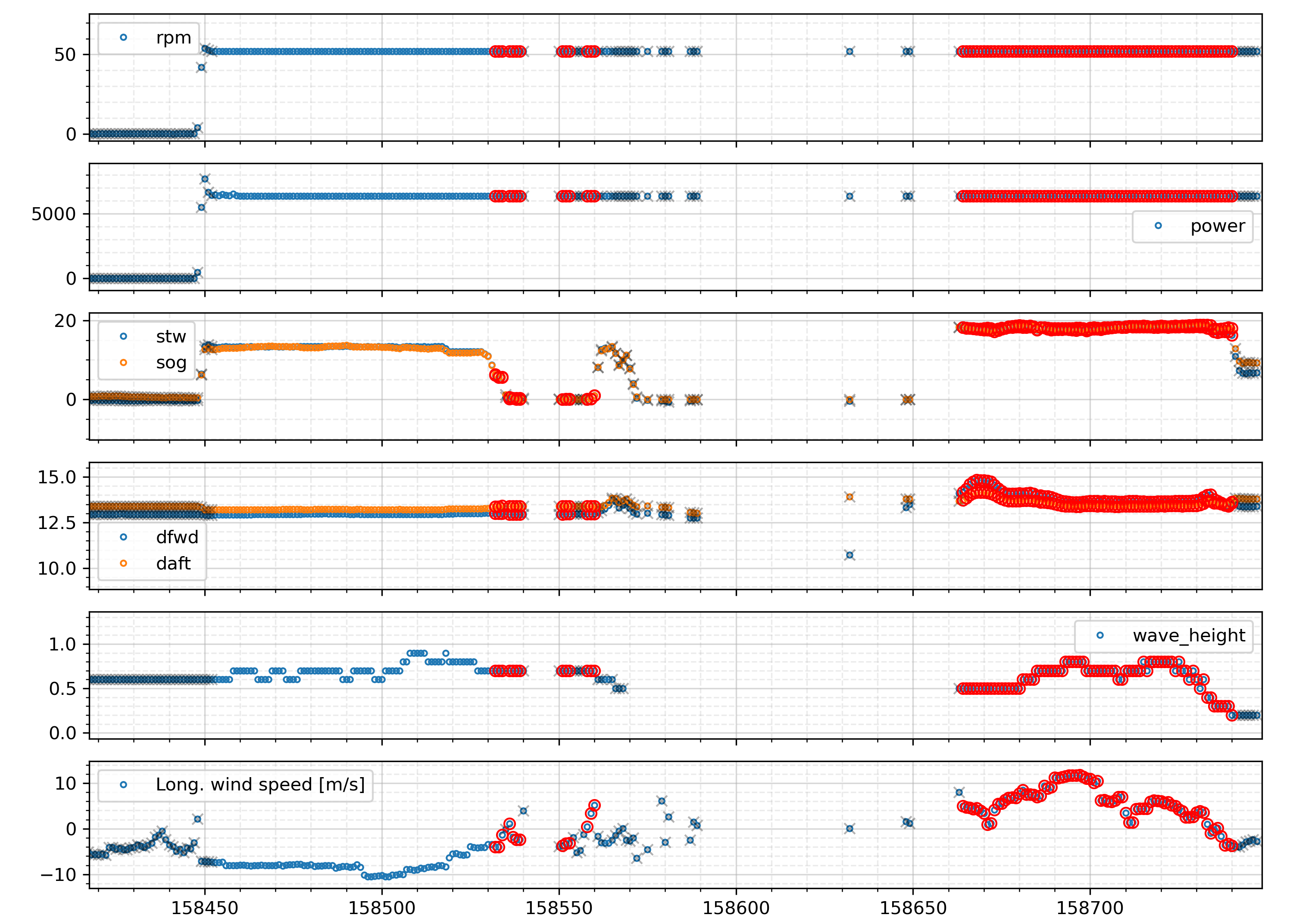}
\caption{Time-series.} \label{fig:corrDefyingOutliersTS}
\end{subfigure}
\caption{Correlation-defying outliers marked with red circles.} \label{fig:corrDefyingOutliers}
\end{figure}

As discussed in section \ref{sec:outliers}, the outliers can be divided into two broad categories: (a) Contextual outliers, and (b) Correlation-defying outliers. The contextual outliers can be identified and resolved by the methods presented as well as demonstrated by \citet{Dalheim2020DataPrep}, and for correlation-defying outliers, methods like Principal Component Analysis (PCA) and autoencoders can be used. Figure \ref{fig:corrDefyingOutliers} shows the in-service data samples recorded onboard a ship. The data here is already filtered-out for quasi-steady assumption, explained above, and contextual outliers, according to the methods suggested by \citet{Dalheim2020DataPrep}. Thus, the samples highlighted by red circles (around 6.4 MW shaft power in figure \ref{fig:corrDefyingOutliersSP}) can be classified as correlation-defying outliers. The time-series plot (shown in figure \ref{fig:corrDefyingOutliersTS}) clearly indicates that the detected outliers have faulty measurements for the speed-through-water (stw) and speed-over-ground (sog), defying the correlation between these variables and the rest. It is also quite surprising to notice that the same fault occurs in both the speed measurements at the same time, considering that they are probably obtained from different sensors. 

\section{Conclusion} \label{sec:conclusion}

The quality of data is very important in estimating the performance of a ship. In this study, a streamlined semi-automatic data processing framework is developed for ship performance analysis. The data processing framework can be used to process data from several different sources, like onboard recorded in-service data, AIS data and noon reports. These three data sources are discussed here in detail along with their inherent problems and associated examples. It is here recommended to use the onboard recorded in-service data for ship performance monitoring over the other data sources, as it is considered more reliable due its consistent and higher sampling rate. Moreover, the AIS data and noon reports lacks some of the critical variables required for ship performance analysis, and they are also susceptible to human error, as some of the data variables recorded here are manually logged by the ship's crew. Nevertheless, all three data sources are known to have several problems and should be processed carefully for any further analysis. 

The data processing framework, presented in the current work, is designed to address and resolve most of the problems found in the above three data sources. It is first recommended to divide the data into trips so that further processing can be performed in a more systematic manner. A simple logic to divide the data into individual trips is outlined here if the port call information is not available. The weather hindcast (metocean) data is considered as an important supplementary information, which can be used for data validation and estimating environmental loads experienced by the ship. A simple algorithm to effectively interpolate the hindcast data at a specific time and location of a ship is presented within the data processing framework. The problem of erroneous draft measurements, caused due to the Venturi effect, is discussed in detail as well as simple interpolation is recommended to fix these measurements. A more complex case, where the draft or trim is voluntarily adjusted during the voyage without reducing the vessel speed, is also presented here. Such a case cannot be resolved with simple interpolation, and therefore, an alternate method is suggested for the same problem. 

Choosing the most suitable methods for estimating resistance components may also be critical for ship performance analysis. It is, therefore, recommended to carry out some validation checks to find the most suitable methods before adopting them into practice. Such validation checks should be done, wherever possible, using the data obtained from the ship while in-service rather than just using the sea trial or model test results. Data cleaning and outlier detection is also considered an important step for processing the data. Since cleaning the data requires selecting a subset of features relevant for the analysis, it is recommended to perform this as the last step of the data processing framework, and some part of it should be reiterated before carrying out a new type of analysis. The presented data processing framework can be systematically and efficiently adopted to process the datasets for ship performance analysis. Moreover, the various data processing methods or steps mentioned here can also be used elsewhere to process the time-series data from ships or similar sources, which can be used further for a variety of tasks.

% Main text
% \section{}\label{}

% Use the style of numbering in square brackets.
% If nothing is used, default style will be taken.
%\begin{enumerate}[a)]
%\item 
%\item 
%\item 
%\end{enumerate}  

%\begin{itemize}
%\item 
%\item 
%\item 
%\end{itemize}  

%\begin{description}
%\item[]
%\item[] 
%\item[] 
%\end{description}  

% Figure
% \begin{figure}[<options>]
% 	\centering
% 		\includegraphics[<options>]{}
% 	  \caption{}\label{fig1}
% \end{figure}

% \begin{table}[<options>]
% \begin{tabular*}{\tblwidth}{@{}LL@{}}
% \toprule
% \midrule
%  & \\
%  & \\
%  & \\
%  & \\
% \bottomrule
% \end{tabular*}
% \end{table}

% Uncomment and use as the case may be
%\begin{theorem} 
%\end{theorem}

% Uncomment and use as the case may be
%\begin{lemma} 
%\end{lemma}

\appendix

% \input{6.Appendix.tex}

% \section{}\label{}

\printcredits

\bibliographystyle{cas-model2-names}

\bibliography{bibliography}



\end{document}